# Unified Theory of Thermal Quenching in Inorganic Phosphors

Mahdi Amachraa[a†], Zhenbin Wang[b†], Chi Chen[b], Shruti Hariyani[c], Hanmei Tang[a], Jakoah Brgoch[c], Shyue Ping Ong[b*]

[a]Materials Science and Engineering Program, University of California San Diego, 9500 Gilman Dr, Mail Code 0418, La Jolla, CA 92093-0448, United States

[b]Department of NanoEngineering, University of California San Diego, 9500 Gilman Dr, Mail Code 0448, La Jolla, CA 92093-0448, United States

[c]Department of Chemistry, University of Houston, College of Natural Sciences and Mathematics, Fleming Building, Room 112, Houston, Texas 77204-5008

**Abstract**: We unify two prevailing theories of thermal quenching (TQ) in rare-earth-activated inorganic phosphors – the cross-over and thermal ionization mechanisms – into a single predictive model. Crucially, we have developed computable descriptors for local activator environment stability from *ab initio* molecular dynamics simulations to predict TQ under the cross-over mechanism, which can be augmented by a band gap calculation to account for thermal ionization. The resulting TQ model predicts the experimental TQ in 29 known phosphors to within ~ 3-8%. Finally, we have developed an efficient topological approach to rapidly screen vast chemical spaces for the discovery of novel, thermally robust phosphors.

## INTRODUCTION

Lighting accounts for approximately 15% of global energy consumption and 5% of $CO_2$ emissions.[1] Solid-state lighting (SSL), based on phosphor-converted light-emitting diodes (pc-LEDs), is ~ 10x more efficient than traditional lighting and therefore offers a huge opportunity to achieve substantial energy savings and $CO_2$ reductions. A critical component in pc-LEDs is the rare-earth substituted inorganic phosphor, which down-converts the near-ultraviolet or blue LED emission to longer wavelengths. The phosphors currently employed in these bulbs comprise an inorganic host material, such as an oxide or nitride, activated with a rare-earth ion that is typically $Ce^{3+}$ or $Eu^{2+}$. The prototypical phosphor used in a majority of devices is yttrium aluminum garnet, $Y_3Al_5O_{12}$, activated with $Ce^{3+}$ (YAG:$Ce^{3+}$) to produce a bright yellow emission. Exciting this phosphor when it is coated on top of a blue LED chip produces a broad emission across the visible spectrum light, appearing as white light.

Developing phosphors with high quantum efficiency and excellent thermal stability is a long-standing quest within the SSL community. Emission loss with increasing temperature (also known as thermal quenching or TQ) is of particular importance in next-generation SSL, where high-power LEDs and laser-based LEDs are becoming more common. In combination with smaller device packaging, the heat generated in these devices can negatively influence the optical output, in particular from the phosphor.[2] The TQ of a phosphor is experimentally determined by taking the ratio between the integrated light intensity emitted at operating temperature (~ 423–473 K) and the integrated intensity of light emitted at room temperature. Commercial phosphors, such as YAG:$Ce^{3+}$, have a TQ of less than 10%, meaning a majority of the emission intensity is maintained at high temperature, whereas other phosphors can be entirely thermally quenched (TQ = 100%) at high temperature. It is therefore unsurprising that extensive efforts have been devoted to the investigation of the TQ mechanism in phosphors.[3–6]

Two dominant theories have been proposed to explain the TQ behavior in $Ce^{3+}$ and $Eu^{2+}$-activated phosphors. In the 1960s, Blasse *et al.* proposed that TQ is the result of the non-radiative relaxation of electrons from the excited state to the ground state.[7,8] This "cross-over" mechanism is represented schematically using the configurational coordinate diagram in Figure 1, where the energy difference between the relaxed excited state and the cross-over point $E_a^{co}$ determines the activation barrier for this process. This theory is one of the reasons why there is a search for structurally rigid phosphor hosts. The fundamental assumption here is that a more rigid host prevents access to soft phonon modes, reducing the probability of non-radiative relaxation from the excited configuration to the ground state configuration. However, subsequent experiments have found many violations of this relationship; for example, the $Ca_7Mg(SiO_4)_4$:$Eu^{2+}$, $CaMgSi_2O_6$:$Eu^{2+}$, and $Sr_6M_2Al_4O_{15}$:$Eu^{2+}$ (M = Y, Lu, Sc) phosphors all suffer from TQ despite their seemingly rigid crystal structures, as estimated from their comparatively high Debye temperatures.[9,10] The second theory, attributed to Dorenbos, posits that TQ is due to the thermal excitation of the excited 5*d* electron of $Ce^{3+}$/$Eu^{2+}$ to the conduction band of the host;[6] the activation barrier of this thermal ionization process ($E_a^i$ in Figure 1) determines the TQ of a phosphor, and this barrier is in turn correlated to the band gap of the phosphor host.[11] Other TQ pathways have been suggested as well, e.g., the temperature dependence of the conduction band minimum is expected to lower $E_a^i$ and induce TQ,[3] although the two competing mechanisms ("cross-over" mechanism and thermal ionization process) dominate in the loss of luminescence as a function of temperature in $Ce^{3+}$/$Eu^{2+}$ doped phosphors.

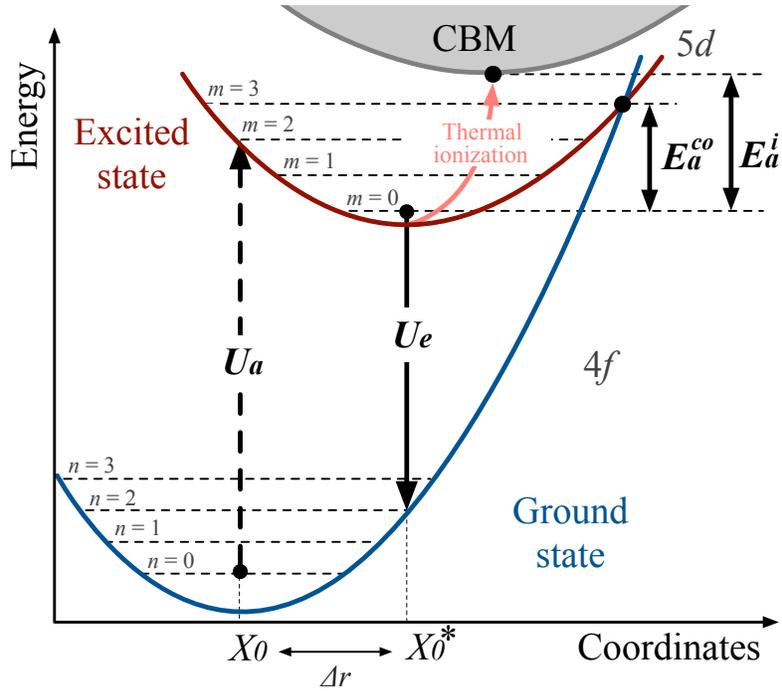

**Figure 1**. Configurational coordinate diagram for the activator in a phosphor. Excitation is allowed from the vibrational level n = 0 of the ground state to the excited state and results in the absorption energy $U_a$. The relaxation of the system from the vibrational level m = 0 to the ground state results in the emission energy $U_e$. The displacement $\Delta r = X_0^* - X_0$ is the polyhedron average bond length difference between the excited and ground state of the activator. In the cross-over model, thermal quenching results from the non-radiative relaxation of an excited electron to the ground state when the temperature is high enough to overcome the activation energy $E_a^{co}$. In the Dorenbos thermal ionization model, thermal quenching results from the promotion of excited electrons to the conduction band minimum if the temperature is high enough to overcome the activation energy $E_a^i$.

In this work, we develop a unified theory of TQ in phosphors by considering the competition between the two dominant TQ mechanisms. It should be noted that all other forms of quenching, thermal or otherwise, are ignored. The focus of this work is specifically on TQ in $Ce^{3+}$ and $Eu^{2+}$-activated oxide phosphors, which are the most common rare-earth ions employed in phosphors and operate on the highly efficient $4f^n5d^0 \rightarrow 4f^{n-1}5d^1$ transition. Using *ab initio* molecular dynamics simulations (AIMD), we establish that the local environment stability around the rare-earth ion as a function of temperature is a robust, transferable descriptor for TQ under the cross-over mechanism when the host band gap is sufficiently large. If the host band gap is small, however, the thermal ionization mechanism competes with the cross-over mechanism. We, therefore, develop a model accounting for both quenching mechanisms to predict the TQ of known phosphors. A total of 29 oxide phosphors with experimentally-measured TQ values were selected to construct the models. Information on these phosphors is summarized in Table S1. Our results show that the combination of AIMD simulations and band gap ($E_g$) calculations provides a clear relationship between a series of computed descriptors and a phosphor's thermal quenching behavior. Furthermore, we propose a novel topological descriptor based on the Voronoi analysis that can be used to rapidly screen for low TQ phosphors without expensive AIMD simulations, thereby allowing the discovery of new thermally robust inorganic phosphors.

## METHODS

**Density Functional Theory (DFT) calculations.** All DFT calculations were performed using the Vienna ab initio simulation package (VASP) within the projector-augmented wave method.[12,13] The exchange-correlation interaction was described using the Perdew-Berke-Ernzerhof (PBE)[14] generalized gradient approximation (GGA) functional with the Hubbard $U$ extension. In general, the parameters used are similar to those used in the Materials Project[15], with a plane wave energy cutoff of 520 eV and a $k$-point density at least of 100 per Å$^{-3}$. A $U$ value of 2.5 eV was used for the $4f$ orbitals in Eu and Ce following previous works.[12,16] All structures were fully relaxed with energies, and forces converged to within $10^{-5}$ eV and 0.01 eV/Å, respectively.

To construct the $Ce^{3+}/Eu^{2+}$-activated phosphors, $Ce^{3+}/Eu^{2+}$ was doped into all compatible symmetrically distinct sites in supercells of the host crystal with lattice parameters of at least 10 Å in each direction. The lowest energy $Ce^{3+}/Eu^{2+}$-doped structure was then used for subsequent analysis and AIMD simulations. All crystal structure and data analysis were carried out using the Python Materials Genomics (pymatgen) package.[17]

**Ab initio molecular dynamics (AIMD) simulations.** AIMD simulations were carried out on the supercell models of $Ce^{3+}/Eu^{2+}$-activated phosphors in the NVT ensemble at 300 K and 500 K with a Nose-Hoover thermostat.[18] The simulation cell was fixed at the final 0 K relaxed cell parameters for each phosphor. For reasons of computational efficiency, the AIMD simulations were non-spin-polarized, and a minimal Γ-centered $1 \times 1 \times 1$ $k$-point mesh and a time step of 2 fs were adopted. Given that the main output extracted from the AIMD simulations is the local environment fluctuations, we expect this approximation to be reasonable.

**Host Debye temperature calculations.** The Debye temperatures of all host compounds were calculated using the quasi-harmonic model given by:

$$\Theta_D = \frac{h}{2\pi k_B}\left(6\pi^2 V^{\frac{1}{2}} n\right) f(v) \sqrt{\frac{B}{M}}$$

where, $V, n, f(v), V$ and $M$ are the unit cell volume, the number of atoms in the unit cell, a scaling function in terms of Poisson's ratio $v$, the bulk modulus and the molar mass, respectively; $h$ and $k_B$ refer to the Planck's constant, and the Boltzmann constant, respectively.[19]

**Activator local environment determination.** The local environments of the activator ions were computed using the algorithm of Waroquiers *et al.* implemented in pymatgen.[20] Additionally, Hoppe's effective coordination number (ECoN) was utilized to determine bond weight based on geometry, leading to a chemically guided local environment algorithm.[21] Based on the polyhedron geometry with the highest weight, a unique cut-off radius or maximum distance factor (MDF) was determined and used to determine the local environment distribution changes during AIMD simulations. The MDF $\alpha$ was computed as follows:

$$\alpha = \frac{l^x}{l^x_{min}}$$

where $l^x$ and $l^x_{min}$ are the ligand-activator bond length and the smallest activator-ligand bond length, respectively, and the ECoN values were computed as follows:

$$\text{ECoN} = \sum \exp\left(1 - \left(\frac{l_i}{l_{avg}}\right)^6\right)$$

$$l_{avg} = \frac{\sum\left(l_i \exp\left(1 - \left(\frac{l_i}{l_{min}}\right)^6\right)\right)}{\sum\left(\exp\left(1 - \left(\frac{l_i}{l_{min}}\right)^6\right)\right)} \qquad (1)$$

$l_i$, $l_{min}$, and $l_{avg}$ are the ligand-activator bond length, the lowest ligand-activator bond length, and the weighted average bond length, respectively.

The Voronoi representation computes a polar transformation of an activator's nearest neighbors into a Cartesian frame and requires an additional angular parameter:

$$\gamma = \frac{\theta^x}{\theta^x_{max}}$$

$\theta^x_{max}$ and $\theta^x$ are the widest angle to uniquely define a given set of ligands and the angle at which the local environment is considered, respectively.

**Average cation electronegativity and centroid shift.** The centroid shift, $\varepsilon_c$, was expressed by Dorenbos[22] as follows:

$$\varepsilon_c = \frac{1.44 \times 10^{17} \alpha_{sp} N_a}{R_{eff}^6}$$

where, $\alpha_{sp}$ is the effective spectroscopic polarizability, which for oxides is given as[23] $\alpha_{sp} = 0.4 + \frac{4.6}{\chi_{av}^2}$, where, $\chi_{av}$ is defined as the average cation electronegativity;[24] $N_a$ is the number of anions in the first surrounding shell and $R_{eff}$ is the effective bond length estimated as:

$$R_{eff} = \frac{1}{N_a}\sum_{i=1}^{N_a}(R_i - 0.6\Delta R)$$

where $R_i$ are the bond lengths between the activator ($Ce^{3+}/Eu^{2+}$) site to the ligands in the host structure, and $\Delta R$ accounts for the difference in ionic radius between the activator and the cation it substitutes for. In this work, $R_{eff}$ and $N_a$ are replaced with Hoppe's $l_{avg}$ (equation 1) and ECoN, respectively, computed using the DFT-relaxed phosphor structures containing the rare-earth ion, leading to the following expression for the centroid shift:

$$\varepsilon_c^{DFT} = \frac{1.44 \times 10^{17} \alpha_{sp} \text{ECoN}}{l_{avg}^6} \qquad (2)$$

**Estimation of the crystal field splitting.** The point charge electrostatic model (PCEM)[25] provides the most straightforward geometrically-driven approach to describe $\varepsilon_{cfs}$ in terms of crystal field parameters denoted by $B_q^k$; where $k$ and $q$ values depend on the point group symmetry of the centroid's polyhedron. The $B_q^k$ parameters are defined as the product of two independent components: a radial part $f(R)$ and an angular part $\Theta_q^k$. Although the PCEM is not well suited for computing accurate $\varepsilon_{cfs}$ values, it has been previously established that there is a positive correlation between the fourth-rank angular parameter ($\Theta_q^4$), $R_{eff}^{-2}$, and $\varepsilon_{cfs}$ in $Ce^{3+}$-doped phosphors in the octahedral, cubic, tricapped trigonal prismatic, and cuboctahedron geometries, given by:

$$\varepsilon_{cfs}(R) = \beta_{poly} R_{eff}^{-2}$$

where $\beta_{poly}$ is the shape factor and is dependent on the polyhedron geometry and correlated to the angular part $\Theta_q^4$, while $R_{eff}^{-2}$ accounts for the radial part of $\varepsilon_{cfs}$.[22] The general PCEM expression for the $B_0^2$, and $B_0^4$ crystal field parameters have the form:

$$B_0^2 = f(R)\,\Theta_0^2 = Ze^2\frac{\langle r^2\rangle}{R_{eff}^3}\left[p - \frac{n}{2} + m(3\cos^2(\theta_{pr}) - 1)\right]$$

$$B_0^4 = f(R)\Theta_0^4 = Ze^2\frac{\langle r^4\rangle}{R_{eff}^5}\left[p + \frac{3n}{8} + \frac{m}{4}(35\cos(\theta_{pr})^4 - 30\cos(\theta_{pr})^2 + 3)\right]$$

where, $Ze$ is the effective charge on the ligands (oxygen in this study) separated by a distance $R_{eff}$ from the centroid ($Ce^{3+}/Eu^{2+}$), $r^2$ or $r^4$ is the expectation value of the radial distance to the of the $5d$ orbital from the nucleus second or fourth power and its value is assumed to be constant, $p$ is the number of axial ligands, $n$ is the number of equatorial ligands, $m$ is the number of ligands in the base plane of the prism, and $\theta_{pr}$ is the prismatic angle between the $2m$ prismatic ligands and the $m$-fold rotational axis. Three types of activator geometries are found within the $Ce^{3+}$-doped phosphors considered in this study, namely, cubic, octahedral and trigonal prismatic with point groups $O_h$, $O_h$, and $D_{3h}$, respectively (see Figure 2). Both the $O_h$ and $D_{3h}$ point groups can be defined under a 3-fold rotation axis. The cubic and octahedral geometries share the same 3-fold rotational axis with the following $\Theta_0^2$, and $\Theta_0^4$ values:

$$\Theta_0^2(oct) = 0;\ \Theta_0^4(oct) = -2.33$$
$$\Theta_0^2(cub) = 0;\ \Theta_0^4(cub) = 2.07$$

$\Theta_0^4(oct)$ has a negative value, i.e., the $e_{2g}$ orbitals in an octahedral environment experience repulsive forces, while $\Theta_0^4(cub)$ has a positive value, and the $e_{2g}$ orbitals in a cubic environment experience high attractive forces. For the trigonal prismatic with a $D_{3h}$ symmetry, ($\theta_{pr} = 49°, m = 3, CN = 6$) the PCEM values for $\Theta_0^2$ and $\Theta_0^4$ can be extracted as follows:

$$\Theta_0^2 = m[3\cos^2(\theta_{pr}) - 1] = 0.87$$
$$\Theta_0^4 = m\frac{m}{4}[(35\cos(\theta_{pr})^4 - 30\cos(\theta_{pr})^2 + 3)] = -2.57$$

The polyhedron geometries around the $Eu^{2+}$ considered in the study generally exhibit complex geometries defined under different rotational axis. In this work, we utilize Hoppe's effective coordination to identify the highest weight bonds, which are likely to have the highest effect on the $\varepsilon_{cfs}$, followed by Waroquiers et al.'s algorithm to identify all possible symmetries an activator's polyhedron can have with respect to small local atomic displacements (see Figure S1 for a flowchart). The 18 $Eu^{2+}$ local environments in this study are then classified into four distinct geometries: $D_{3h}$ trigonal prism, $O_h$ octahedron, $C_{2v}$ mono-capped or bi-capped trigonal prism (1CTP or 2CTP), and $I_h$ icosahedron. In this work, we assume that the crystal field splitting of a mono-capped trigonal prism is comparable to other crystal field splitting when defined under a 3-fold rotation axis, as suggested by previous results.[26] The bi-capped trigonal prism is approximated as a mono-capped trigonal prism. Therefore, for the mono/bi-capped trigonal prism with a $C_{2v}$ symmetry, ($\theta_{pr} \sim 47°, p = 0, n = 1, m = 3, CN = 7$) the PCEM values for $\Theta_0^2$ and $\Theta_0^4$ can be extracted as follows:

$$\Theta_0^2(1CTP) = 0.69;\ \Theta_0^4(1CTP) = -2.16$$

The icosahedron with $I_h$ symmetry possesses a 5-fold rotation axis, which is not compatible with its crystallographic space group; hence only distorted icosahedra are allowed to occur in real crystals, where the second and fourth-rank parameters are equal to zero in both the 5-fold and 3-fold rotation axis.[25] The local environment approximations considered in this work are essential to defining all crystal-field parameters under a unique quantization axis (3-fold), ensuring the direct comparison of all computed parameters.

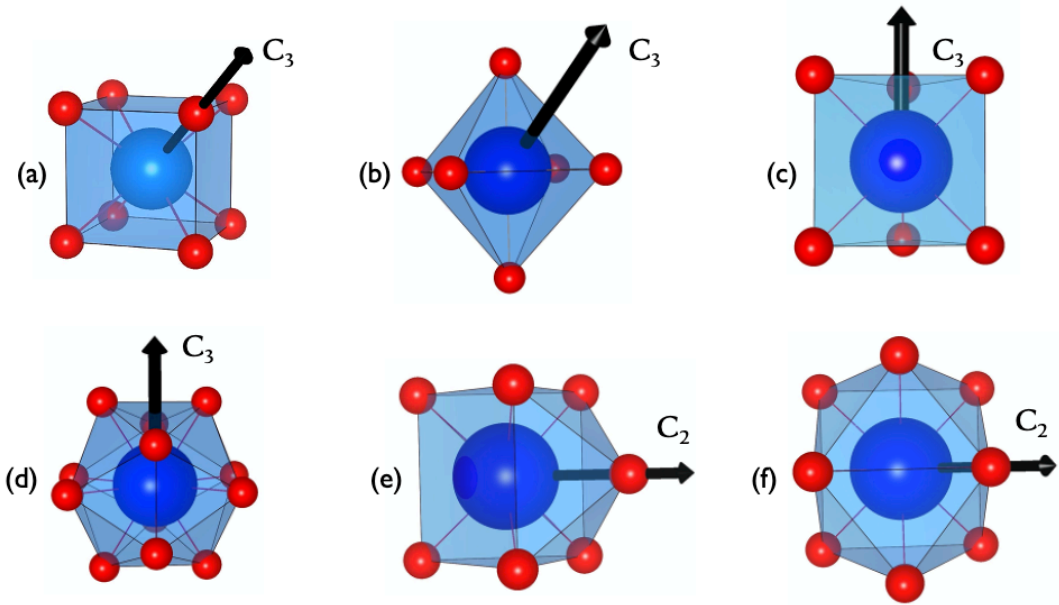

**Figure 2.** Local environments in the 29 $Ce^{3+}/Eu^{2+}$-doped phosphors: (a) cube ($O_h$), (b) octahedron ($O_h$), (c) trigonal prism ($D_{3h}$), (d) icosahedron ($I_h$) (e) mono-capped trigonal prism ($C_{2v}$), and (f) bi-capped trigonal prism ($C_{2v}$). $C_3$ and $C_2$ refers to the rotational axis. It should be noted that the local environments shown here are the idealized coordination polyhedra (without distortion). Actual local environments in the phosphors may be distorted from these idealized shapes.

## RESULTS AND DISCUSSION

### Relationship between Thermal Quenching and Local Environment Rigidity

AIMD simulations were performed on the 29 phosphors at room temperature (300 K) as well as 500 K to represent an upper-temperature limit experienced by the phosphor during lamp operation.[27] An activator environment distribution (AED) at both temperatures was constructed for each phosphor from the AIMD trajectories. The AED is derived by determining the number of simulation timesteps that the activator has a particular coordination number (CN) using the algorithm developed by Waroquiers *et al.*,[20] normalized across the total number of timesteps. Figure 3 presents the AED for three $Ce^{3+}$ and three $Eu^{2+}$-activated phosphors with high (> 80%), intermediate (40-60%) and low (< 20%) TQ. Similar plots for the remaining 23 phosphors are provided in Figures S2 and S3. Comparing the AEDs between 300 K and 500 K show that that thermally robust (small TQ) phosphors generally exhibit minimal change in the AED as a function of temperature whereas thermally quenched (high TQ) phosphors show substantial shifts in the AED with temperature. For example, the $Ce^{3+}$ in YAG:$Ce^{3+}$ (TQ = 9%) is primarily eight-fold coordinated with oxygen and its CN remains stable at 500 K. In contrast, the $Ce^{3+}$ in $Ba_2Y_5B_5O_{17}$:$Ce^{3+}$ (TQ = 71%) has a distribution of CN of 7-8 at 300 K and exhibits an obvious shift to lower CNs, including seven-fold and six-fold coordination environments, at 500 K. The same trend is observed for $Eu^{2+}$-activated phosphors in Figure 3b.

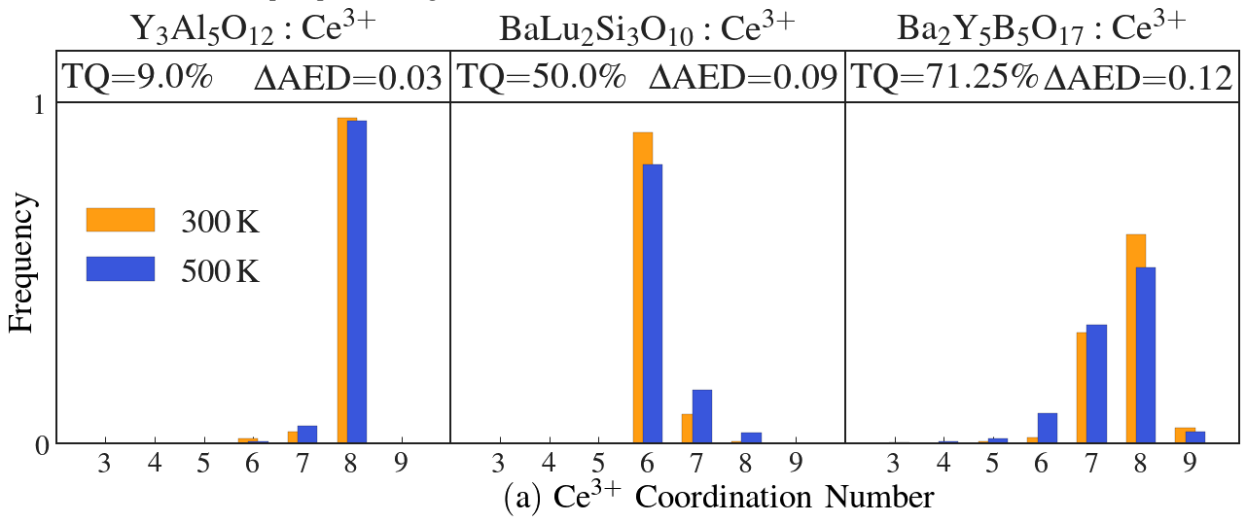

(a) $Ce^{3+}$ Coordination Number

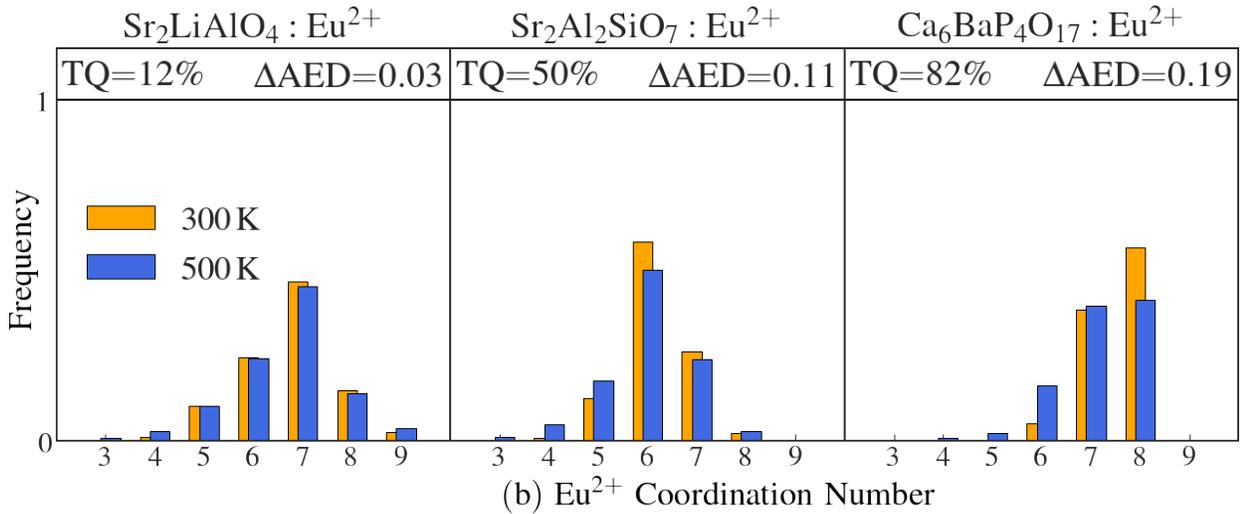

**Figure 3**. Activator environment distribution (AED). The computed AED of (a) $Ce^{3+}$ and (b) $Eu^{2+}$ at 300 K and 500 K in three hosts with distinct TQ behaviors. The experimental TQ values were obtained from references [28–36]

To quantify the shift in the AED from 300 K to 500 K, we define ΔAED as the Euclidean distance between AEDs at 300 K and 500 K in Equation 3,

$$\Delta\text{AED} = \sqrt{\sum_{x=2}^{12}(\omega_{CN=x}^{300K} - \omega_{CN=x}^{500K})^2} \quad (3)$$

where $\omega_{CN=x}^{300K}$ and $\omega_{CN=x}^{500K}$ are the normalized CN frequencies at 300 K and 500 K, respectively, and the CN ranges from 2 to 12. Figure 4 shows the experimentally measured TQ ($TQ_{exp}$) plotted against the computed ΔAED for $Ce^{3+}$ and $Eu^{2+}$-activated phosphors. We find there is an approximately-linear positive correlation between TQ and ΔAED for both $Ce^{3+}$ and $Eu^{2+}$-activated oxide phosphors. A least-squares fitting for the one-parameter expression TQ = KΔAED, where K is a constant, yields $R^2$ values of 0.89 and 0.62 for $Ce^{3+}$ and $Eu^{2+}$-activated phosphors, respectively, with reasonable root mean square errors (RMSEs) of 11.6% and 14.3%, respectively. In contrast, the DFT-calculated Debye temperature ($\Theta_D$) yields a weaker correlation against $TQ_{exp}$ (see Figure S4) with $R^2$ values of 0.14 and 0.12 for $Ce^{3+}$ and $Eu^{2+}$-activated phosphors, respectively, and corresponding RMSEs of 32% and 26%, respectively. There is some correlation between ΔAED and host structural rigidity, as suggested by the high Debye temperatures, the coordinated local displacements within a crystal environment are more directly related to the low ΔAED, making it more reliable compared to a global descriptor like structural rigidity. These observations support our hypothesis that ΔAED is an effective descriptor to probe the depth of the potential energy surface (PES) in the cross-over model, where a higher ΔAED implies a shallower PES in the ground and excited states, which results in a lower $E_a^{co}$ and higher TQ.

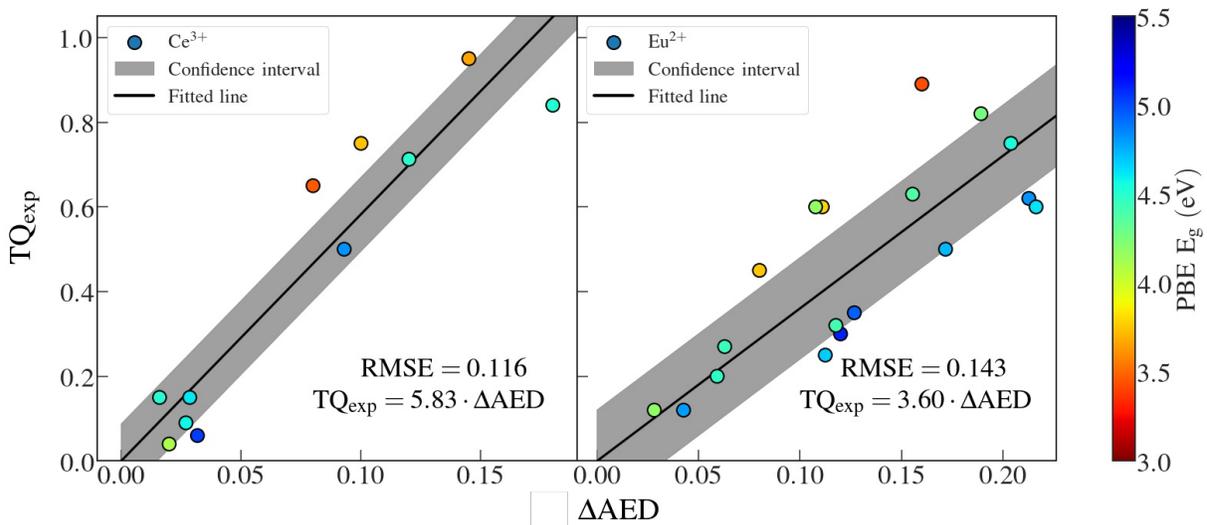

**Figure 4**. Local environment rigidity. Plot of experimental thermal quenching ($TQ_{exp}$) against the change in the activator environment distribution (ΔAED) for $Ce^{3+}$ (left), $Eu^{2+}$ (right) activated phosphors. The marker colors are based on their calculated band gap using the Perdew-Burke-Ernzerhof (PBE) functional (see color bar). The root mean square error (RMSE) for each model is defined as the square root of the average squared difference between the computed ΔAED and the experimental TQ ($TQ_{exp}$), i.e. RMSE = $\sqrt{\frac{1}{N}\sum_{i=1}^{N}(TQ_{exp}^i - K\Delta AED^i)^2}$.

**Accounting for Thermal Ionization**

A closer examination of Figure 4 further reveals that phosphors with large host band gaps ($E_g$) lie below the regression line, and those with small $E_g$ lie above the line. This observation suggests that $E_g$, which is related to TQ under the Dorenbos single-barrier quenching model, also plays a critical role for some materials.

Figure 5 shows a schematic of the energy levels of a $Ce^{3+}/Eu^{2+}$ activator in a host. It can be seen that the Dorenbos thermal ionization barrier is given by:

$$E_a^i(A^{x+}, H) = E_g - \left[E(A^{x+}, \text{free}) - \varepsilon_c(A^{x+}, H) - \frac{\varepsilon_{cfs}(A^{x+}, H)}{r(H)} + \varepsilon_s(A^{x+}, \text{free})\right] - E(4f^n - VBM) \quad (4)$$

where, $E_g$ is the band gap of the host, $E(A^{x+}, \text{free})$ is the fixed centroid position for the free $A^{x+}$ ion ($E(Ce^{3+}, \text{free}) = 6.118$ eV and $E(Eu^{2+}, \text{free}) = 4.216$ eV),[37] $\varepsilon_c(A^{x+}, H)$ is the centroid shift for $A^{x+}$ in the host H, $\varepsilon_{cfs}(A^{x+}, H)$ is the crystal-field splitting experienced in the host H, $r(H)$ expresses the ratio between the crystal field splitting and the crystal field shift (typical values range between 1.7 and 2.4 and are geometry dependent), $\varepsilon_s(A^{x+}, \text{free})$ is the energy difference between the centroid position and the lowest 5$d$ level of the free $A^{x+}$ ion and is also a constant, and $E(4f^n - VBM)$ is the energy difference between the $4f^n$ of $A^{x+}$ and the valence band maximum (VBM) of the host H. The binding energy $E(4f^n)$ for $Ce^{3+}$ has been found to be relatively constant for garnets and perovskites and will be assumed constant for the 11 $Ce^{3+}$-doped hosts considered in the study, while $E(4f^n)$ for $Eu^{2+}$ has an energy fluctuation of ~ 0.38 eV in oxides depending on the host H.[22] The $r(H)$ is assumed 5/2 for 6-coordinated geometries, and 5/3 for 8-coordinated geometries.[25] The $\varepsilon_c(Ce^{3+}, H)$ was evaluated using the ligand polarization model where the spectroscopic polarizability is quantitatively correlated to $\chi_{av}$ as defined in the Methods section. Finally, the susceptibility of the 5$d$ energy levels induces an additional redshift, namely, the crystal field splitting $\varepsilon_{cfs}(A^{x+}, H)$, which can be determined by the geometry around the $Ce^{3+}/Eu^{2+}$ activators as outlined in the Methods section. For brevity, we will drop the explicit functional dependence of the variable henceforth.

Based on equation 4, we, therefore, express the thermal ionization barrier as follows:

$$E_a^i = A E_g^{DFT} + B \varepsilon_c^{DFT} + \frac{1}{r}\left(C \frac{\Theta_0^2}{l_{avg}^3} + D \frac{\Theta_0^4}{l_{avg}^5}\right) + E \quad (5)$$

where $\varepsilon_{cfs}$ is expressed as the sum of both second and fourth-rank crystal-field parameters:

$$\varepsilon_{cfs} = C \frac{\Theta_0^2}{l_{avg}^3} + D \frac{\Theta_0^4}{l_{avg}^5}$$

where *A*, *B*, *C*, *D* and *E* are fitted constants. *A* accounts for the well-known systematic underestimation of $E_g$ by the PBE functional,[38] while *B* accounts for systematic errors in the estimation of the centroid shift from the DFT lattice parameters (Equations 1 and 2). *C* and *D* constants account for the mutual importance of the second and fourth-rank crystal-field parameters, as well as the refactoring of the computed $\varepsilon_{cfs}$ energies with respect to the computed $E_g^{DFT}$ and $\varepsilon_c^{DFT}$.

**Figure 5.** Energy level diagram for $Ce^{3+}/Eu^{2+}$ activators (denoted by $A^{x+}$) in a host material (denoted by H). The $\varepsilon_c(A^{x+}, H)$, $\varepsilon_{cfs}(A^{x+}, H)$, $D(A^{x+}, H)$ and $E(A^{x+}, \text{free})$ refer to the centroid shift, crystal-field splitting, the redshift of the activators' $4f^{n-1}5d^1$ level in a host H, respectively, and the energy of the first $f \to d$ transition the free ion $A^{x+}$. The VBM and the CBM indicate the valence band maximum and the conduction band minimum, respectively. Note that the energy of the VBM is the energy referential and is set to 0 eV.

Considering the cross-over model and the thermal ionization model for quenching are two independent sources of TQ, both are likely occurring simultaneously in some phosphor systems. Therefore, we combine ΔAED with the Dorenbos expression for TQ[6] under the thermal ionization model (detailed derivations are provided in Supporting Information) to arrive at the following formula of $TQ_{pred}$:

$$TQ = 1 - (1 - K\Delta AED)\frac{1 + \Gamma e^{\frac{-E_a^i}{k_B T_1}}}{1 + \Gamma e^{\frac{-E_a^i}{k_B T_2}}} \quad (6)$$

where, $k_B$ is the Boltzmann constant, $T_1$ (300 K) is the initial temperature in Kelvin, $T_2$ is the final temperature of quenching, which in these calculations is 500 K, $\Gamma$ is defined as the ratio of the attempt rate for thermal quenching ($\Gamma_0$) and the radiative decay rate of the $5d$ state ($\Gamma_v$)[6,28], and $E_a^i$ is the barrier for thermal ionization under the Dorenbos model.

**Table 1.** Optimized coefficients from non-linear least-squared minimization of unified TQ model for $Ce^{3+}$ and $Eu^{2+}$ activated hosts.

| Activator | A | B | C | D | E | K | K' | β |
|---|---|---|---|---|---|---|---|---|
| $Ce^{3+}$ | 0.15 | 0.15 | 0.13 | 0.037 | 0.180 eV | 4.05 | 0.27 | 2.75 |
| $Eu^{2+}$ | 0.19 | 0.012 | 0.21 | 0.41 | 0.049 eV | 3.22 | 0.77 | 0.86 |

The optimal values of $K$, $A$, $B$, $C$, $D$, and $E$ (as defined in equation 5) were determined by performing a non-linear least-square minimization of the multi-variable predicted TQ from Equation 6 with the experimentally-observed TQ of the 29 phosphors (see Table S1) and are tabulated in Table 1. The $Ce^{3+}$ and $Eu^{2+}$-activated phosphors are expected to have different $A$, $B$, $C$, $D$, and $E$ values because of their large difference in energy gap between the $4f$ ground state and the $5d$ excited state (6.2 eV and 4.2 eV for free $Ce^{3+}$ and $Eu^{2+}$ ions, respectively) and different ionic radii.[37] Most importantly, the band gap factor ($A$=0.15 and 0.19) for $Ce^{3+}$ and $Eu^{2+}$ is expected to be the same since both are computed using the same functional. Figure 6 plots the predicted $TQ_{pred}$ using the optimized equation 6 against the experimental $TQ_{exp}$. The RMSE for $Ce^{3+}$ and $Eu^{2+}$ are 3.1% and 7.6%, respectively, which are a significant improvement over the model using ΔAED alone. Further validation of our unified TQ model can be seen in the fact that the predicted $E_a^i$ (Table S1) are in good agreement with experimental thermal ionization energies. For example, the predicted $E_a^i$ of Sr$_2$MgSi$_2$O$_7$:Eu$^{2+}$, SrSc$_2$O$_4$:Eu$^{2+}$, Y$_3$Al$_5$O$_{12}$:Ce$^{3+}$, Lu$_3$Al$_5$O$_{12}$:Ce$^{3+}$ and K$_3$YSi$_2$O$_7$:Ce$^{3+}$ are 0.87 eV, 0.71 eV, 0.88 eV, 0.87 eV and , 0.54 eV respectively, and the corresponding experimental thermal ionization energies are 0.9 eV, 0.56 eV, 0.77 eV, 0.75 eV and 0.48 eV, respectively.[22,28,39,40,41] Note, that our predicted $E_a^i$ for Y$_3$Al$_5$O$_{12}$:Ce$^{3+}$ is higher than the predicted value for Lu$_3$Al$_5$O$_{12}$:Ce$^{3+}$ as shown in the vacuum referred binding energies diagram.[22]

Previous findings from Dorenbos have suggested that the redshift $D$ (the sum of the centroid shift and the crystal splitting shift) of the 5$d$ state of the Ce$^{3+}$ activator is correlated to the one of Eu$^{2+}$ ion when inserted within the same host/site as follows:

$$D(\text{Eu}^{2+}, \text{H}) = 0.64 D(\text{Ce}^{3+}, \text{H}) - 0.233 \text{ eV} \pm 0.15$$

Equivalently, photoluminescence properties such as emission energy and activation energies of Ce$^{3+}$-doped phosphors can help assess the photoluminescence of Eu$^{2+}$ when inserted within the same host/site. To further substantiate the validity of our fitted *A*, *B*, *C*, *D*, and *E*, we predict the $E_a^i$ of K$_3$YSi$_2$O$_7$:Eu$^{2+}$ from our fitted results of K$_3$YSi$_2$O$_7$:Ce$^{3+}$ by utilizing Dorenbos's semi-empirical relationship. The predicted $E_a^i$ of Eu$^{2+}$ doped in K$_3$YSi$_2$O$_7$ from the fitted parameters of Ce$^{3+}$-doped compounds are 0.65 eV, while the experimentally-determined values are 0.63 eV, which are in excellent agreement.

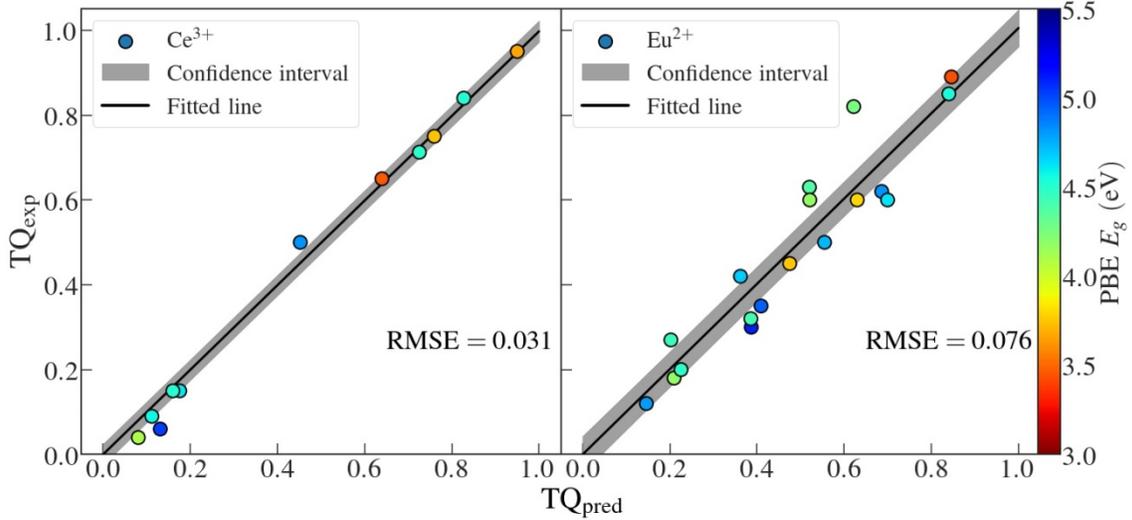

**Figure 6.** Plot of TQ$_{\text{exp}}$ against the predicted TQ (TQ$_{\text{pred}}$) from the unified TQ model (equation 6) for Ce$^{3+}$ (left) and Eu$^{2+}$ (right)-activated hosts. Here, RMSE is computed as follows: RMSE = $\sqrt{\frac{1}{N}\sum_{i=1}^{N}(\text{TQ}_{\text{exp}}^i - \text{TQ}_{\text{pred}}^i)^2}$

The unified TQ equation (Equation 6) can be understood intuitively by considering the competition between the cross-over and thermal ionization mechanisms in Figure 1. The primary loss of emission with temperature increase will come from the mechanism with the lowest barrier $E_a$. When the band gap of the host is large, $E_a^i \gg E_a^{co}$, the cross-over mechanism dominates. ΔAED then describes the depth of the potential energy surface and, hence, TQ. When the band gap of the host is sufficiently small, both mechanisms compete, and ΔAED, $E_g$, $E_g$, $\varepsilon_c$, and $\varepsilon_{cfs}$ are required to describe TQ under a dual-barrier quenching model. Unlike the semi-empirical thermal ionization relationship, the proposed unified TQ model allows for the first time to develop a universal comparison amongst all hosts' high temperature photoluminescence behavior *via* robust descriptors (ΔAED, $E_g$, $\varepsilon_c$, $\varepsilon_{cfs}$) obtained in a *pre-hoc* manner. Moreover, the $E_a^i$ derived from Dorenbos's semi-empirical relationship is a function of T$_{1/2}$ (the temperature at which the intensity reaches half of the initial intensity), where the latter varies for every host. Consequently, all previously obtained $E_a^i$ mentioned in the literature reflect a thermal quenching occurring across different temperatures ranges, and hence cross-comparisons of $E_a^i$ was impossible and impractical until now. Finally, while the proposed dual-barrier thermal quenching offers a comprehensive overview of the TQ mechanism, its applicability cannot directly be extended to phosphors with multiple luminescent centers, as the AED and electronic properties are expected to differ from the pristine structure. Additionally, phosphors known to have photoluminescence compensation mechanisms driven by the formation of thermally activated defect levels are also not within the scope of the dual-barrier quenching model;[42] however, with enough data, an additive model accounting for the electronic contributions of defects can be derived.

**Intrinsic Topological Descriptor of Activator's Stability.** Finally, we demonstrate how a modified version of the unified TQ model can be used to computationally screen for low TQ phosphors. While $E_g^{DFT}$, $\varepsilon_c$, and $\varepsilon_{cfs}$ can be obtained using relatively inexpensive ground-state DFT computations, ΔAED requires expensive AIMD simulations for activated phosphors, where relatively large supercells of the host crystal are required to simulate the experimental low activator concentration. It is, therefore, desirable to establish an alternative descriptor for local environment stability. The AED at 300 K and 500 K reveals that activators are susceptible to endure local polyhedron changes due to mutual oscillations of the centroid/activator and ligands/oxygen in the first shell. The computed mean-square displacements (MSD) of the activator and oxygen ligands increase from 300 K to 500 K, suggesting that the overall occupied 3-dimensional space by the activator and ligands increases with temperature. For example, the MSD behavior of $Ce^{3+}$ and oxygen ligands in the first shell in $Ca_3Sc_2Si_3O_{12}$ host is shown in Figure S5a; a definite increase in the MSD for both $Ce^{3+}$ and oxygen ligands with respect to temperature is observed. Despite large atomic displacements in $Ca_3Sc_2Si_3O_{12}$ host, the periodic-like oscillations of both $Ce^{3+}$ and oxygen ligands result in a low ΔAED. Conversely, the MSD behavior of $Ce^{3+}$ and oxygen ligands in $Ba_3Y_2B_6O_{15}$ host (Figure S5b) shows a quasi-random oscillation resulting in a larger ΔAED. The same observations can be made for $SrMgAl_{10}O_{17}:Eu^{2+}$ and $Ba_2SiO_4:Eu^{2+}$. (Figure S5c and S5d) Consequently, smaller ΔAED values, are suspected to be correlated to large 3-dimensional spaces around the activator's effective local environment.

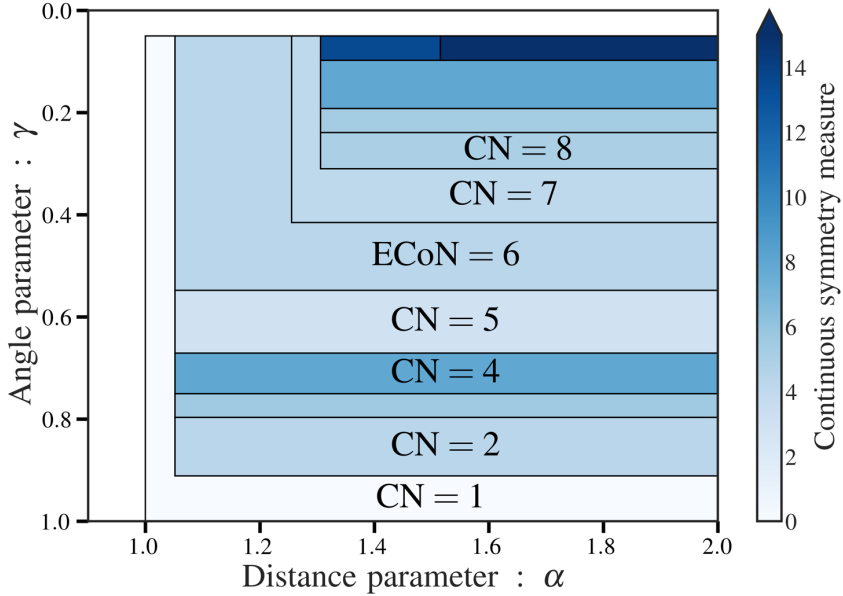

**Figure 7.** The Voronoi grid representation of $Eu^{2+}$ doped in $Sr_2SiO_4$. The Voronoi grid representation has two variables: the angle parameter $\gamma$ and the distance parameter $\alpha$. For a specific range of angle and distance parameters, a unique coordination number is defined. Each Voronoi grid is representative of a set of ligands denoted by the coordination number. The continuous symmetry measure is used to assess the degree of symmetry within different coordination numbers and developed by Pinsky.[43]

A two-dimensional projection of all $Ce^{3+}/Eu^{2+}$ local environments in the 29 hosts is constructed via a Voronoi tessellation-based algorithm. Figure 7 shows a *topological* sensitivity analysis of the local activator environment of $Eu^{2+}$ in $Sr_2SiO_4$. (similar plots for all of the 29 phosphors are provided in the Supporting Information Figure S6 and S7).[17,20] By varying the distance ($\alpha$) and angle ($\gamma$) parameters, the algorithm yields different coordination environments due to changes in the bond weights of the surrounding ligands (see Computational Methodology section for bond weights computations). The coordination environment formed by the highest weight ligands determines the main activator local environment, which in $Sr_2SiO_4:Eu^{2+}$ is ECoN = 6. Our hypothesis is that the larger the normalized area ($\Upsilon$) occupied by the main activator local environment, the less sensitive the activator local environment is to variations in bond distances and angles. In other words, the higher $\Upsilon$, the lower the ΔAED, and the smaller the expected TQ. Therefore, substituting ΔAED with $1 - \Upsilon$ in equation 7, we obtain the following alternative model:

$$TQ = 1 - K'(1-\Upsilon)\frac{1+\Gamma' e^{\frac{-E_a^i}{k_B T_1}}}{1+\Gamma' e^{\frac{-E_a^i}{k_B T_2}}} \quad (7)$$

where $A, B, C, D$ and $E$ for the $E_a^i$ expression in Equation 5 are kept as the optimized values from the non-linear least-squares optimization of equation 6, while $K'$ is refitted. The optimized values of $K'$ are 0.27 and 0.77 for $Ce^{3+}$ and $Eu^{2+}$, respectively (included in Table ). Moreover, the computed $E_a^i$ from Equation 6 is based on the rate of change of ΔAED across all of the 29 compounds, and ΔAED does not scale linearly with respect to $\Upsilon$. Therefore, to compensate for the different rates of ΔAED and $\Upsilon$, the pre-exponential factor is re-defined as $\Gamma' = \beta\Gamma$. The optimized values of $\beta$ for $Ce^{3+}$ and $Eu^{2+}$ were 2.75 and 0.86, respectively (shown in Table 1). Figure 8 plots the $TQ_{exp}$ against $TQ'_{pred}$ as defined by equation 7. The RMSE of $TQ'_{pred}$ for $Ce^{3+}$ and $Eu^{2+}$ is 8.9% and 9.6%, respectively. While this value of RMSE is somewhat higher than the RMSE using Equation 6, the RMSE using Equation 7 is already sufficiently low for $Ce^{3+}/Eu^{2+}$-activated hosts to be used for rapid screening for discovery of low TQ phosphors. The difference in performance between using $(1-\Upsilon)$ and ΔAED, especially for $Ce^{3+}$-activated hosts, can be attributed to two factors. First, some of the considered compounds are known to have multiple symmetrically-distinct doping sites, i.e. $Ba_3Y_2B_6O_{15}:Ce^{3+}$, $Ba_9Lu_2(SiO_4)_6:Ce^{3+}$ and $Ba_9Lu_2(SiO_4)_6:Ce^{3+}$, while the Voronoi area was computed using only the most energetically stable site.[32,44] Second, $(1-\Upsilon)$ is a pure topological descriptor with no consideration of differences in chemical bonding, whereas ΔAED captures subtle relationships between bond distances, bond angles and bond strength in the distribution of activator environments.

Nevertheless, the ability to quickly obtain ϒ values without computationally expensive AIMD calculations makes this approach ideal for materials screening.

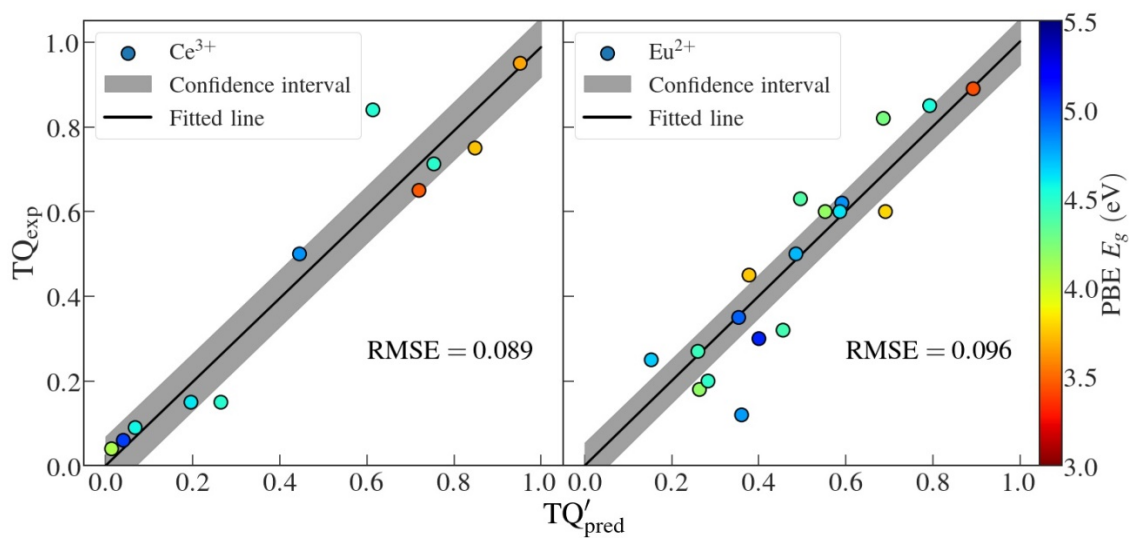

**Figure 8.** Plot of the predicted $TQ_{exp}$ against $TQ'_{pred}$ obtained from the modified TQ model utilizing the topological descriptor in equation 4 for $Ce^{3+}$ (left) and $Eu^{2+}$ (right)-activated phosphors.

## CONCLUSION

In summary, we have developed a unified dual-barrier TQ model by integrating the two prevailing theories – the cross-over and thermal ionization models – for thermal quenching in $Ce^{3+}$ and $Eu^{2+}$-activated phosphors. More critically, we have developed computational approaches to probe thermal quenching in phosphors using this unified TQ model. We establish that local activator environment stability, as measured by the change in activator environment distribution with temperature in AIMD simulations, ΔAED, is the relevant descriptor for TQ under the cross-over mechanism. The computed host band gap ($E_g$), the centroid shift ($\varepsilon_c$), and the approximated crystal field splitting ($\varepsilon_{cfs}$) are descriptors for TQ under the thermal ionization mechanism. This unified dual-barrier thermal quenching model combining ΔAED, $E_g$, $\varepsilon_c$ and $\varepsilon_{cfs}$ predicts the experimentally observed TQ in 29 $Ce^{3+}$ and $Eu^{2+}$-activated phosphors to within an RMSE of 3.1% and 7.6%, respectively. We have also developed an alternative topological descriptor for local environment stability based on Voronoi tessellation that allows for rapid TQ screening of phosphors without expensive AIMD simulations. This work provides crucial insights into the TQ mechanisms in phosphors, and an efficient and reliable way to the discovery of new phosphors with low TQ for next-generation, high power solid-state lighting.

## ASSOCIATED CONTENT

**Supporting Information**

The Supporting Information is available free of charge on the ACS Publications website at DOI: XXX

Derivation of the Unified Dual-barrier Thermal Quenching model, structural and thermal quenching properties of the 29 oxide phosphors, the activator ($Eu^{2+}$/$Ce^{3+}$) environment distribution in all 29 oxide phosphors, and the Voronoi grid representation of the activator's local environment in the 29 oxide phosphors (PDF)

## AUTHOR INFORMATION

**Corresponding Author**

Shyue Ping Ong − Department of NanoEngineering, University of California, San Diego, La Jolla, California 92093-0448, United States; orcid.org/0000-0001-5726-2587; Phone: (858) 534-2668; Email: ongsp@eng.ucsd.edu

**Author Contributions**

[†]These authors contributed equally.
**Notes**
These authors declare no competing financial interest.

## ACKNOWLEDGMENT

This work was primarily supported by the National Science Foundation, Ceramics Program, under grant no. 1911372. S. Hariyani and J. Brgoch also acknowledge funding from the National Science Foundation under grant no. 1911311 as well as the Division of Materials

Research under grant no. 1847701. The computational resources were provided by the Extreme Science and Engineering Discovering Environment (XSEDE) supported by the National Science Foundation under grant no. ACI-1548562, the Triton Super Computer Center (TSCC) at the University of California, San Diego, and the National Energy Research Scientific Computing Center (NERSC).

# Supporting Information

# Unified Theory of Thermal Quenching in Inorganic Phosphors

Mahdi Amachraa[a,†], Zhenbin Wang[b,†], Chi Chen[b], Shruti Hariyani[c], Hanmei Tang[a], Jakoah Brgoch[c], Shyue Ping Ong[b*]

[a]Materials Science and Engineering Program, University of California San Diego, 9500 Gilman Dr, Mail Code 0418, La Jolla, CA 92093-0448, United States

[b]Department of NanoEngineering, University of California San Diego, 9500 Gilman Dr, Mail Code 0448, La Jolla, CA 92093-0448, United States

[c]Department of Chemistry, University of Houston, College of Natural Sciences and Mathematics, Science & Research Building 1, 3507 Cullen Blvd, Room 214, Houston, Texas 77204-5008

**Corresponding Author**

[*] ongsp@eng.ucsd.edu (S. P. Ong)

### Derivation of Unified Thermal Quenching Model

The quantum efficiency (QE) of electronic transitions[1] is given by:

$$QE(T) = \frac{1}{1 + \Gamma e^{\frac{-E_a}{k_B T}}}$$

$$\Gamma = \frac{\Gamma_0}{\Gamma_v}$$

where the constant $\Gamma$ is defined as the ratio of the attempt rate for thermal quenching ($\Gamma_0$) and the radiative decay rate of the 5$d$ state ($\Gamma_v$), $k_B$ is Boltzmann's constant, and $E_a$ is the activation energy for the mechanism of loss of emission. The constant $C$ for $Eu^{2+}$ and $Ce^{3+}$ have previously been derived by Dorenbos as $2.37 \times 10^7$ and as $2.3 \times 10^6$ respectively.[2,3]

The thermal stability (TS) of a phosphor compound is measured as the ratio of the QE between a lower temperature $T_1$ and a higher temperature $T_2$, as follows:

$$TS = \frac{QE(T_2)}{QE(T_1)} = \frac{1 + \Gamma e^{\frac{-E_a}{k_B T_1}}}{1 + \Gamma e^{\frac{-E_a}{k_B T_2}}}$$

Throughout this work, $T_1 = 300$ K and $T_2 = 500$ K. The thermal quenching rate is simply given by:

$$TQ = 1 - TS$$

Assuming that the cross-over and thermal ionization mechanisms operate independently, the overall TS is then given by:

$$TS = TS^{co} TS^i = (1 - K\Delta AED) \frac{1 + \Gamma e^{\frac{-E_a}{k_B T_1}}}{1 + \Gamma e^{\frac{-E_a}{k_B T_2}}}$$

where $\Delta AED$ is the change in local activator environment distribution, the assumption that $TS^{co}$ is a linear function of $\Delta AED$ is used and $K$ is a fitted constant, and $E_a$ being the activation energy barrier under the Dorenbos thermal ionization model can be denoted as $E_a^i$. Therefore, the overall TQ model is given as:

$$TQ = 1 - TS^{co} TS^i = 1 - (1 - K\Delta AED) \frac{1 + \Gamma e^{\frac{-E_a^i}{k_B T_1}}}{1 + \Gamma e^{\frac{-E_a^i}{k_B T_2}}} \tag{S1}$$

where, a thorough derivation of $E_a^i$ is elaborated and included the Main manuscript as follows:

$$E_a^i = AE_g^{DFT} + B\epsilon_c^{DFT} + \frac{1}{r}\left(C\frac{\theta_0^2}{l_{avg}^3} + D\frac{\theta_0^4}{l_{avg}^5}\right) + E \tag{S2}$$

$l_{avg}$ refers to the average bond length as defined in equation 1 in the Main manuscript. Finally, using equation S2 and the results from the AIMD simulations ($\Delta AED$), equation S1 can be solved by fitting the unknown constants $K, A, B, C, D$ and $E$. The constants were determined by performing a least-square minimization of the predicted TQ from equation S1 with the experimentally-observed TQ of the 29 phosphors (given in Table S1). This minimization was carried out using the SciPy package using the BFGS algorithm.[7] All of the computed $\epsilon_c^{DFT}$ are tabulated in Table S1.

Table S1. Structural and thermal quenching properties for 29 oxide phosphors. The TQ data presented here is experimental value collected from corresponding cited references. Act stands for activator. G:CN is the activator's local environment (C=cube, O=octahedron, TP=trigonal prism, 1CTP=mono-capped trigonal prism, 2CTP=bi-capped trigonal prism, I=icosahedron) $E_g$ (unit: eV) is the host band gap calculated using the PBE functional. $\varepsilon_c$ (unit: eV) is the computed centroid shift. ΔAED is the difference in the activator environment distribution (AED) between 300 K and 500 K. The computed Voronoi area is denoted by $Area(\Upsilon)$. $E_a^i$ (unit: eV) is the predicted activation energy. $\Theta_D$(unit: K) is DFT calculated Debye temperature using PBE functional.

| Host material | Space group | Act | G:CN | TQ% | $E_g$ | $\epsilon_c^{DFT}$ | ΔAED | Area(Υ)% | $E_a^i$ | $\Theta_D$ | Refs. |
|---|---|---|---|---|---|---|---|---|---|---|---|
| Lu$_3$Al$_5$O$_{12}$ | $Ia\bar{3}d$ | Ce$^{3+}$ | C:8 | 6 | 5.04 | 1.78 | 0.032 | 88.2 | 0.86 | 605.8 | 3,8,9 |
| Ba$_9$Lu$_2$Si$_6$O$_{24}$ | $R\bar{3}$ | Ce$^{3+}$ | O:6 | 15 | 4.62 | 1.86 | 0.029 | 86.6 | 0.74 | 398.3 | 10 |
| Y$_3$Al$_5$O$_{12}$ | $Ia\bar{3}d$ | Ce$^{3+}$ | C:8 | 9 | 4.58 | 2.25 | 0.027 | 78.4 | 0.88 | 712.8 | 3,11,12 |
| Ca$_3$Sc$_2$Si$_3$O$_{12}$ | $Ia\bar{3}d$ | Ce$^{3+}$ | C:8 | 4 | 4.10 | 1.75 | 0.020 | 80.0 | 1.00 | 672.6 | 13–15 |
| Ba$_9$Y$_2$Si$_6$O$_{24}$ | $R\bar{3}$ | Ce$^{3+}$ | C:8 | 15 | 4.51 | 1.84 | 0.016 | 86.0 | 0.72 | 388.3 | 16 |
| BaLu$_2$Si$_3$O$_{10}$ | $P2_1/m$ | Ce$^{3+}$ | O:6 | 50 | 4.82 | 1.38 | 0.093 | 14.5 | 0.71 | 406.5 | 17 |
| Ba$_2$Y$_5$B$_5$O$_{17}$ | $Pbcn$ | Ce$^{3+}$ | O:6 | 71 | 4.50 | 1.23 | 0.120 | 38.0 | 0.63 | 392.4 | 18 |
| Y$_3$Mg$_2$AlSi$_2$O$_{12}$ | $Ia\bar{3}d$ | Ce$^{3+}$ | C:8 | 75 | 3.94 | 1.68 | 0.100 | 12.0 | 0.61 | 652.0 | 19 |
| Gd$_3$Al$_5$O$_{12}$ | $Ia\bar{3}d$ | Ce$^{3+}$ | C:8 | 65 | 3.46 | 1.78 | 0.080 | 79.1 | 0.63 | 500.9 | 20 |
| Ba$_3$Y$_2$B$_6$O$_{15}$ | $Ia\bar{3}$ | Ce$^{3+}$ | O:6 | 84 | 4.52 | 1.24 | 0.180 | 95.0 | 0.64 | 379.0 | 21 |
| K$_3$YSi$_2$O$_7$ | $P6_3/mmc$ | Ce$^{3+}$ | TP:6 | 95 | 3.67 | 1.46 | 0.145 | 95.0 | 0.54 | 515.2 | 22 |
| SrMgAl$_{10}$O$_{17}$ | $P6_3/mmc$ | Eu$^{2+}$ | TP:6 | 12 | 4.80 | 0.63 | 0.043 | 50.0 | 0.93 | 656.9 | 23,24 |
| KSrPO$_4$ | $Pm$ | Eu$^{2+}$ | 1CTP:7 | 30 | 5.10 | 0.78 | 0.120 | 47.8 | 1.03 | 292.8 | 25,26 |
| KBaPO$_4$ | $Pnm$ | Eu$^{2+}$ | 1CTP:7 | 35 | 4.95 | 0.77 | 0.127 | 54.0 | 1.00 | 332.9 | 26,27 |
| Sr$_2$LiAlO$_4$ | $P2_1/m$ | Eu$^{2+}$ | TP:6 | 12 | 4.19 | 1.25 | 0.028 | 79.0 | 0.82 | 467.5 | 28 |
| BaZrSi$_3$O$_9$ | $P\bar{6}2c$ | Eu$^{2+}$ | O:6 | 25 | 4.68 | 0.63 | 0.112 | 80.0 | 1.25 | 493.1 | 29,30 |
| BaSc$_2$Si$_3$O$_{10}$ | $P2_1/m$ | Eu$^{2+}$ | 2CTP:8 | 50 | 4.73 | 0.90 | 0.172 | 37.0 | 0.96 | 529.2 | 31 |
| Sr$_3$SiO$_5$ | $P4/ncc$ | Eu$^{2+}$ | 2CTP:8 | 45 | 3.76 | 1.53 | 0.080 | 82.0 | 0.78 | 402.7 | 32–34 |
| Ba$_2$MgSi$_2$O$_7$ | $C2/c$ | Eu$^{2+}$ | O:6 | 27 | 4.45 | 0.71 | 0.063 | 66.0 | 1.20 | 386.6 | 35 |
| SrLiPO$_4$ | $P6_3$ | Eu$^{2+}$ | I:12 | 32 | 4.42 | 0.99 | 0.118 | 41.6 | 0.92 | 378.9 | 26 |
| Sr$_2$SiO$_4$ | $Pmcm$ | Eu$^{2+}$ | TP:6 | 63 | 4.38 | 1.09 | 0.156 | 38.0 | 0.88 | 431.7 | 36 |
| BaLu$_2$Si$_3$O$_{10}$ | $P2_1/m$ | Eu$^{2+}$ | 2CTP:8 | 62 | 4.82 | 0.94 | 0.213 | 23.0 | 0.98 | 406.5 | 37 |
| Sr$_2$Al$_2$SiO$_7$ | $P\bar{4}2_1m$ | Eu$^{2+}$ | TP:6 | 60 | 4.20 | 1.08 | 0.111 | 37.0 | 0.75 | 514.6 | 38–40 |
| Ba$_2$SiO$_4$ | $Pmcm$ | Eu$^{2+}$ | 1CTP:7 | 60 | 4.63 | 1.08 | 0.216 | 24.0 | 0.94 | 311.3 | 36 |
| Ca$_6$BaP$_4$O$_{17}$ | $C2/m$ | Eu$^{2+}$ | 2CTP:8 | 82 | 4.26 | 1.04 | 0.189 | 12.0 | 0.88 | 507.8 | 41 |
| Ca$_7$Mg(SiO$_4$)$_4$ | $Pnn2$ | Eu$^{2+}$ | TP:6 | 60 | 4.18 | 1.51 | 0.108 | 34.5 | 0.83 | 601.0 | 42 |
| CaMgSi$_2$O$_6$ | $Pmcm$ | Eu$^{2+}$ | C:8 | 75 | 4.55 | 1.18 | 0.204 | 25.0 | 0.73 | 665.0 | 42 |
| Sr$_2$MgSi$_2$O$_7$ | $P\bar{4}2_1m$ | Eu$^{2+}$ | TP:6 | 20 | 4.49 | 0.64 | 0.059 | 68.0 | 0.87 | 475.7 | 43 |
| SrSc$_2$O$_4$ | $Pnma$ | Eu$^{2+}$ | 2CTP:8 | 89 | 3.44 | 1.67 | 0.160 | 8.38 | 0.71 | 604.7 | 44 |

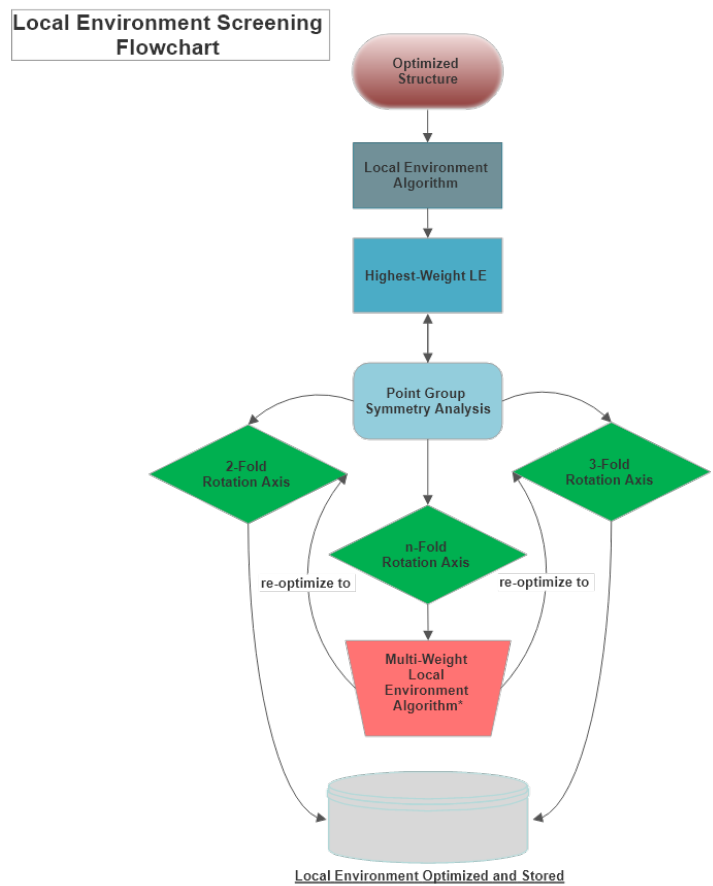

**Figure S1.** Flowchart showing how the local environment (LE) is determined. The highest-weight local environment is obtained by computing the effective coordination number and effective bond weights as developed by Hoppe. The point group symmetry analysis is conducted by the python materials genomics software (pymatgen). Waroquiers's LE algorithm is utilized to compute a multi-weight interrelation by simulating small atomic displacements. Finally, the highest LE is re-optimized to either have a 3-fold or 2-fold rotation axis.

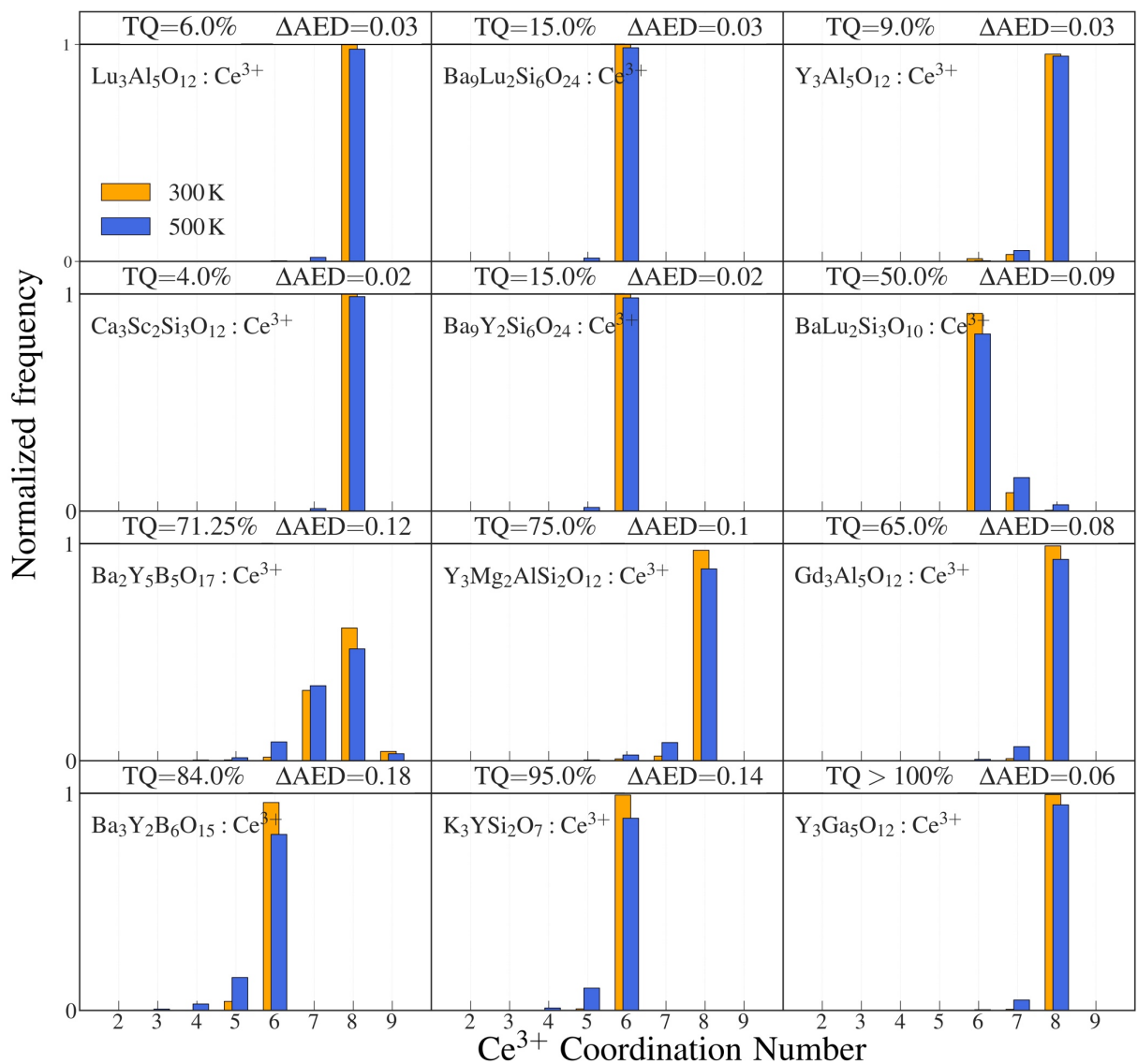

**Figure S2**. The activator environment distribution for $Ce^{3+}$-activated hosts at 300 K and 500 K. Experimental TQ rates as well as the computed ΔAED are shown. Note, the TQ rate of $Y_3Ga_5O_{12}$:$Ce^{3+}$ is not applicable here, as its quantum efficiency is measured to be 0 between 300 K and 500 K. The AED of $Y_3Ga_5O_{12}$:$Ce^{3+}$ are shown for comparison purposes between other garnet structures.

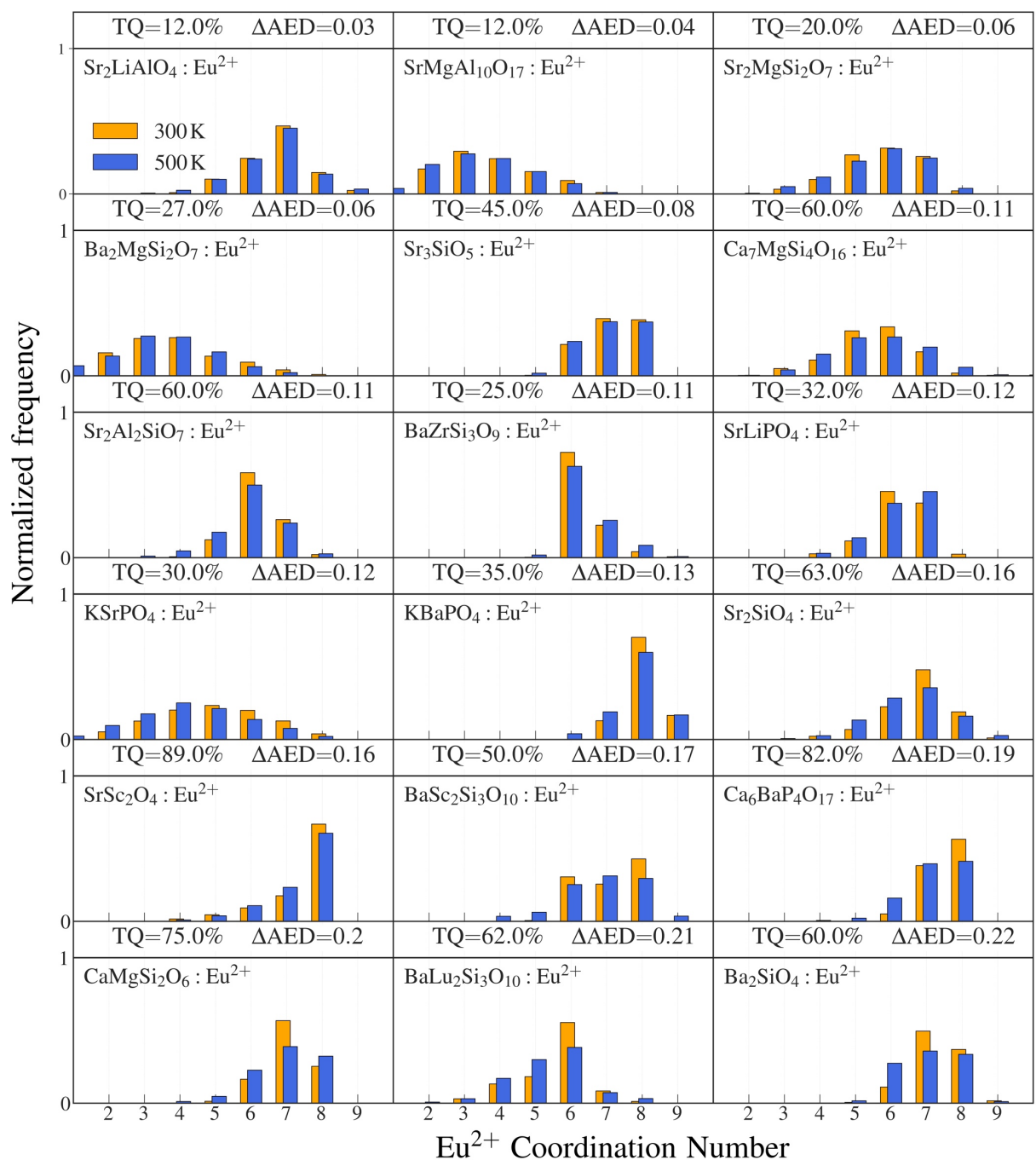

**Figure S3**. The activator environment distribution for $Eu^{2+}$-activated hosts at 300 K and 500 K. Experimental TQ rates as well as the computed ΔAED are shown.

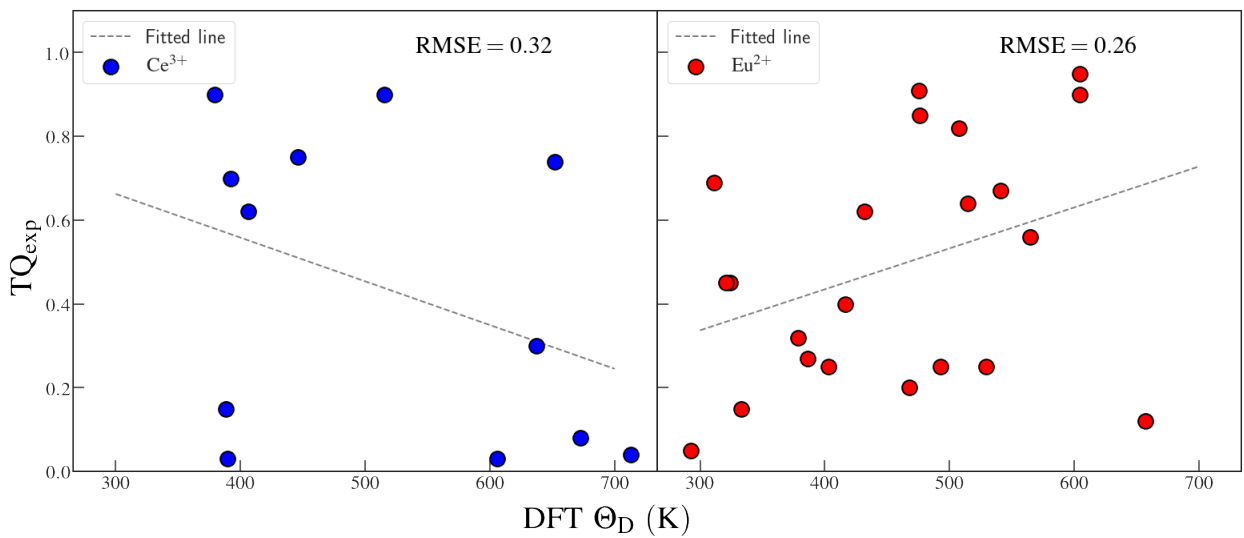

**Figure S4.** Experimentally measured thermal quenching (TQ$_{exp}$) against the DFT-calculated Debye temperature ($\Theta_D$).

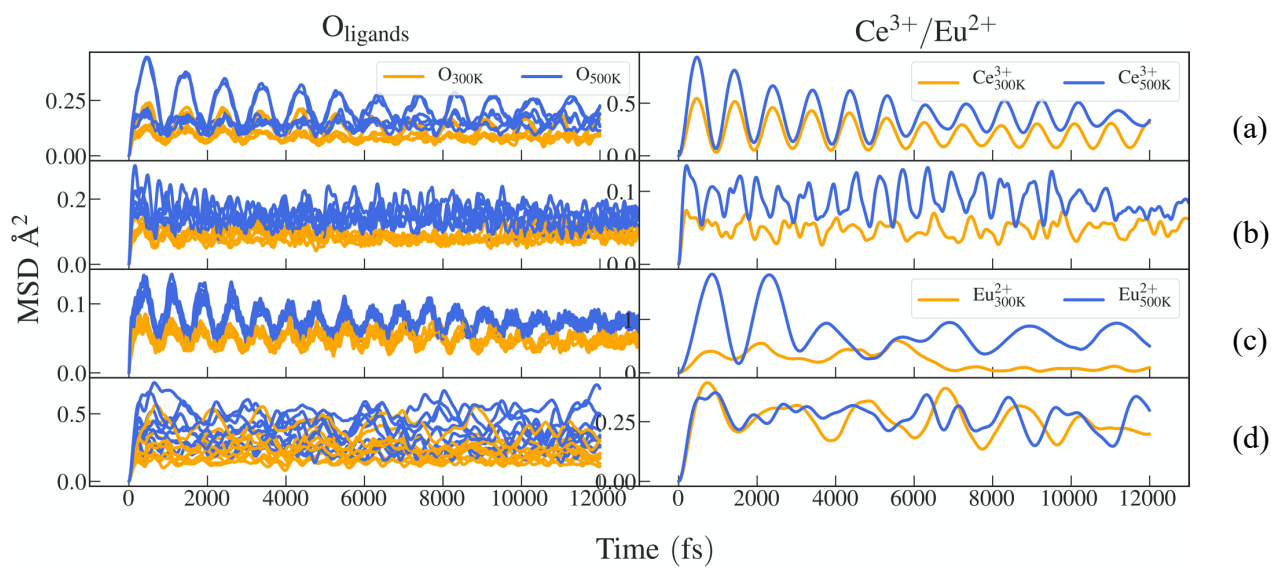

**Figure S5.** The mean-square displacements (MSD) against time (fs) of Ce$^{3+}$/Eu$^{2+}$ and oxygen ligands (O$_{ligands}$) in (a) Ca$_3$Sc$_2$Si$_3$O$_{12}$, (b) Ba$_3$Y$_2$B$_6$O$_{15}$, (c) SrMgAl$_{10}$O$_{17}$, and (d) Ba$_2$SiO$_4$ host. Note, the left-hand plots refer to the MSD behavior of oxygen ligands in the first shell, while the right-hand plots refer to the MSD behavior of Ce$^{3+}$ and Eu$^{2+}$ activator in the doped supercells.

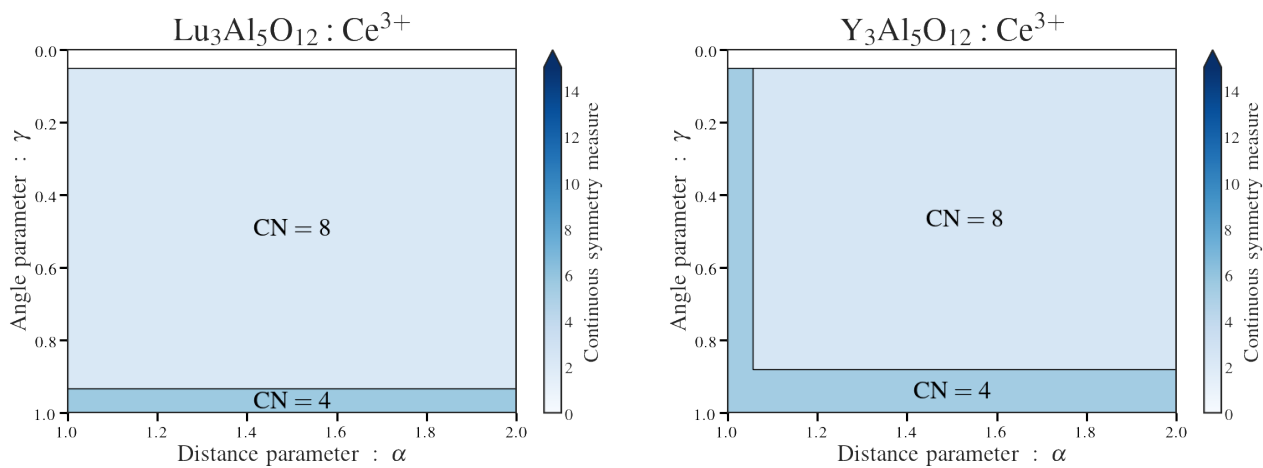

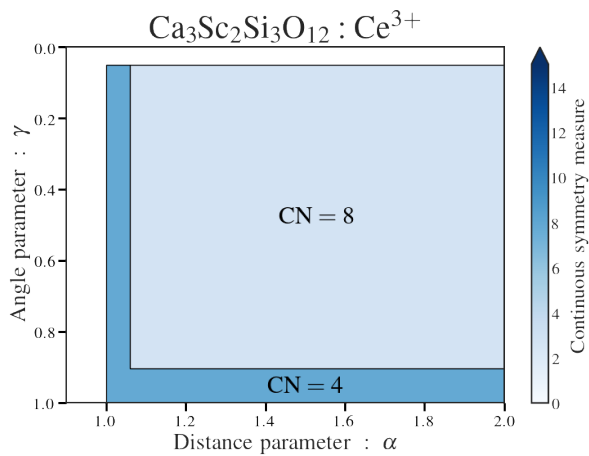
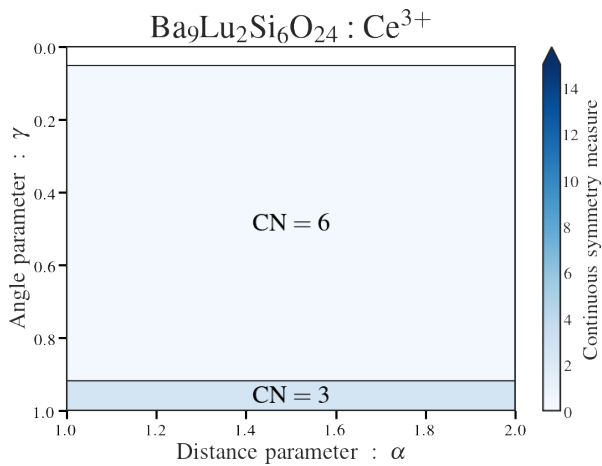
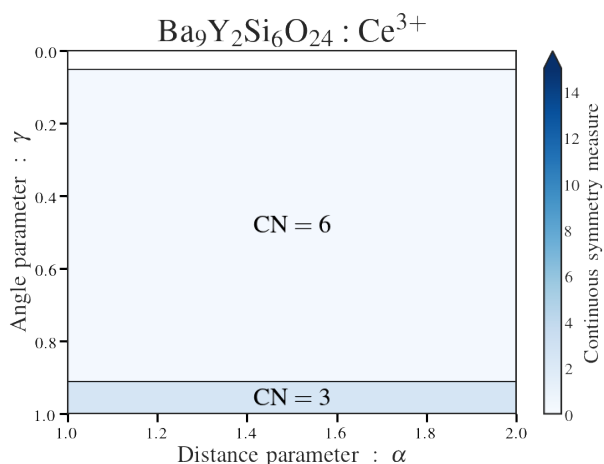
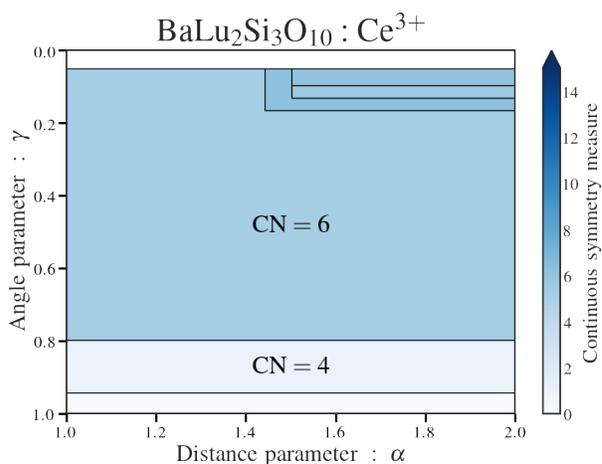
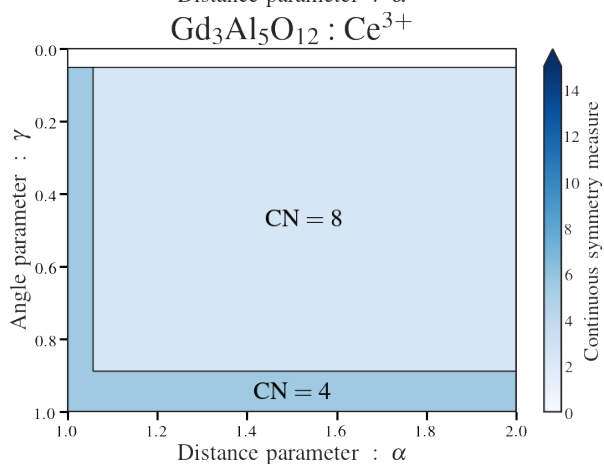
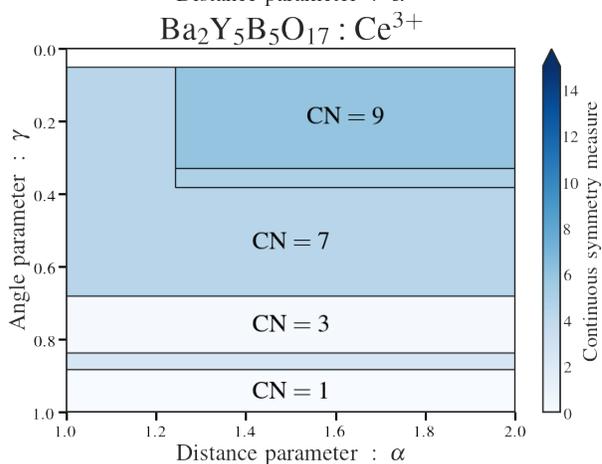
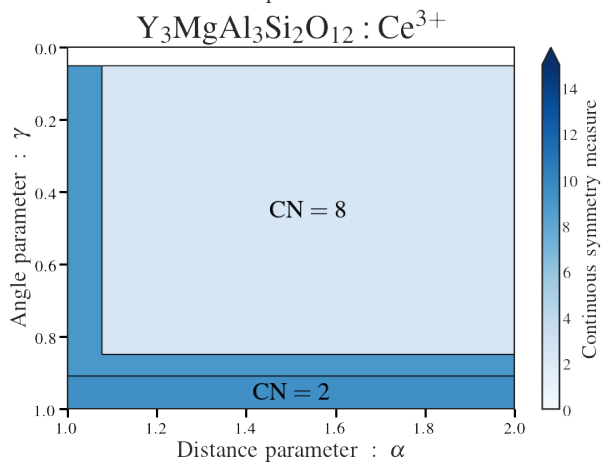
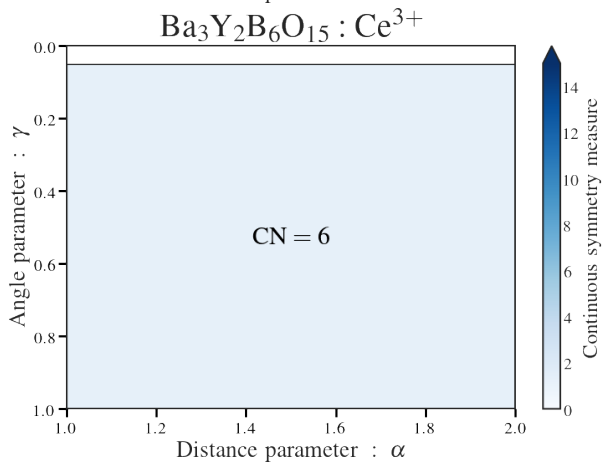

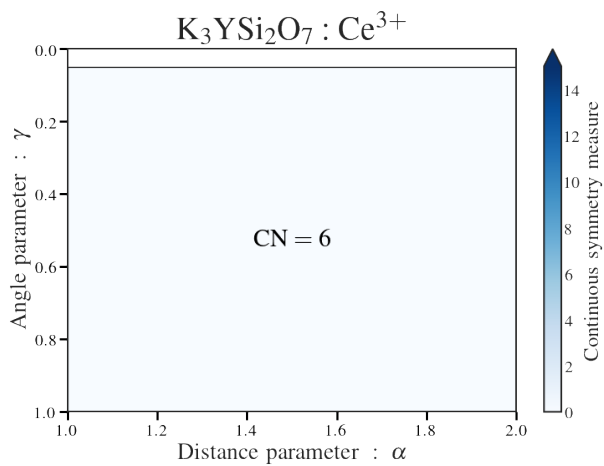
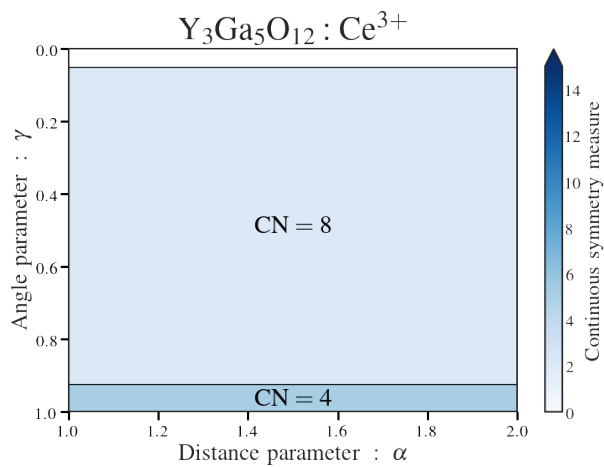

**Figure S6.** The Voronoi grid representation of $Ce^{3+}$ local environment of all $Ce^{3+}$-activated hosts in Table S1. The coordination number (CN) in red reflects the environment considered to compute the Voronoi area.

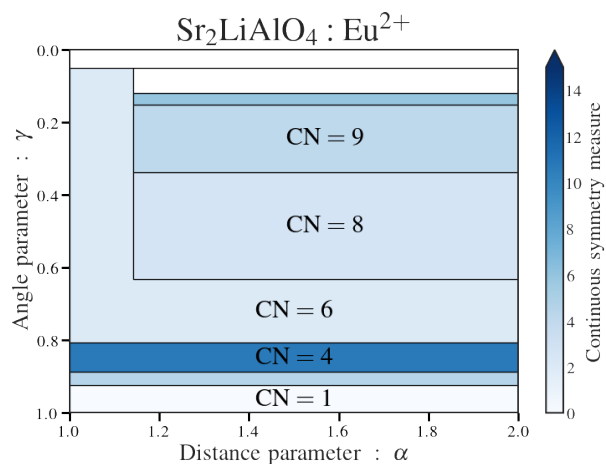
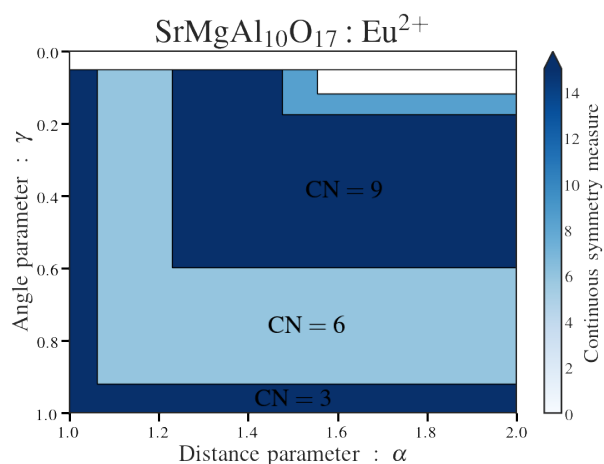
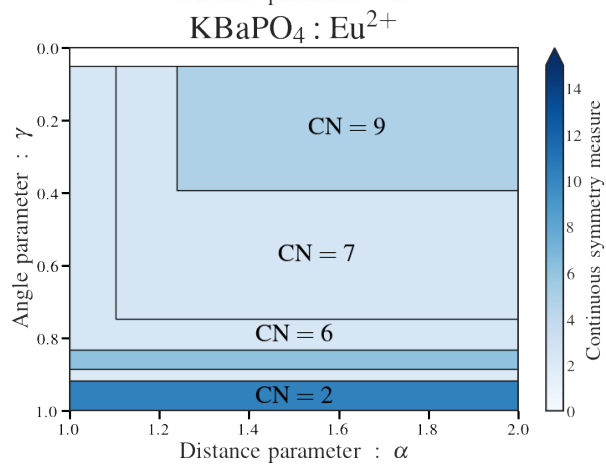
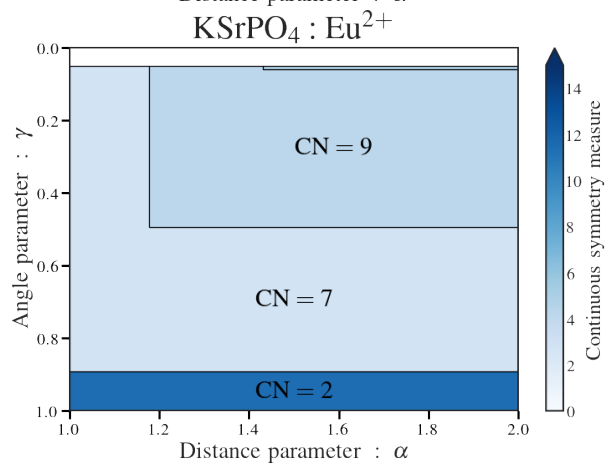

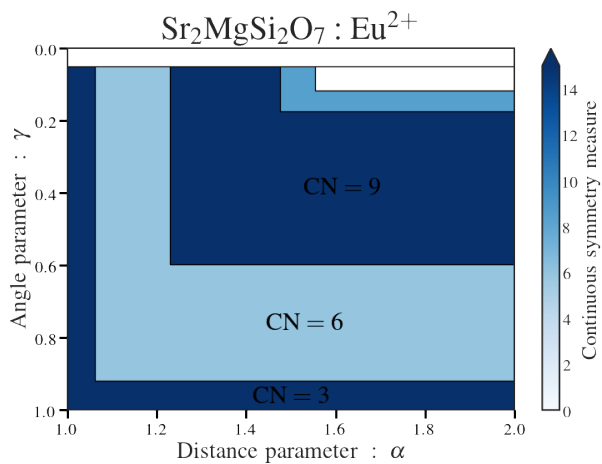
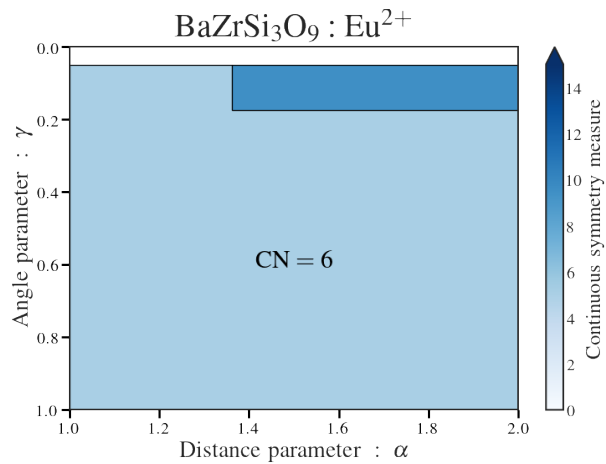
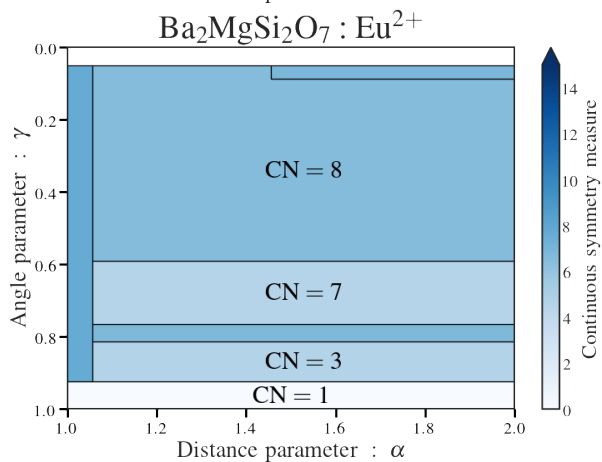
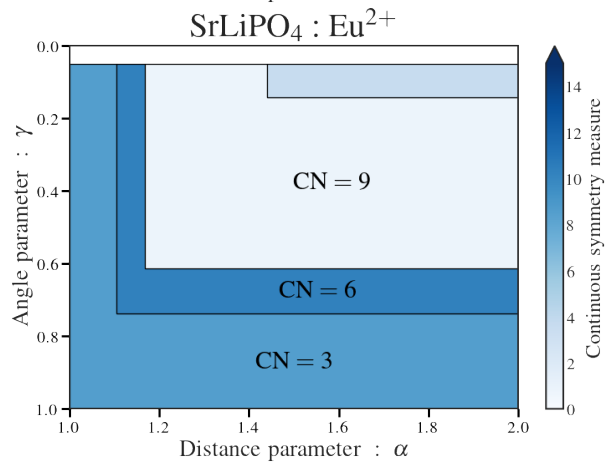
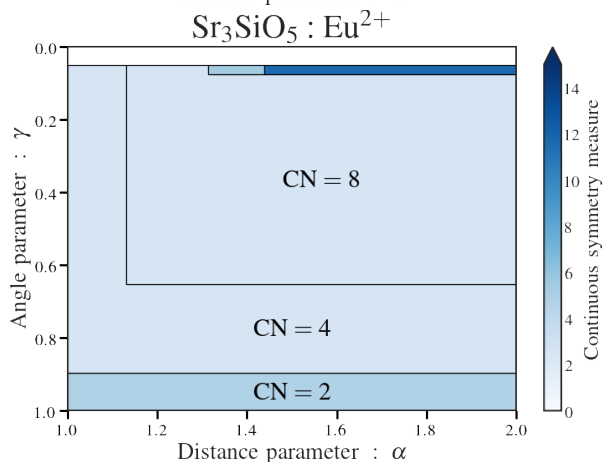
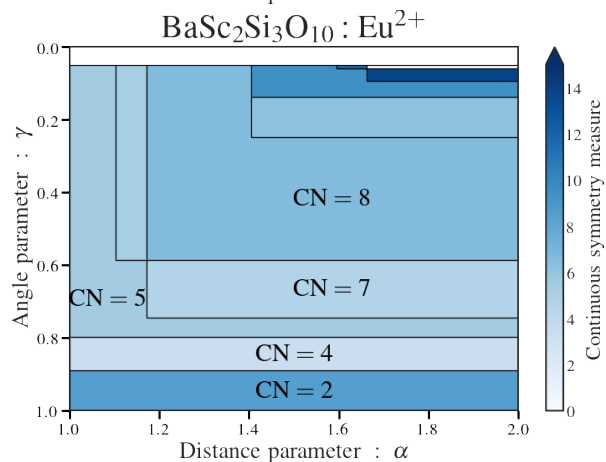
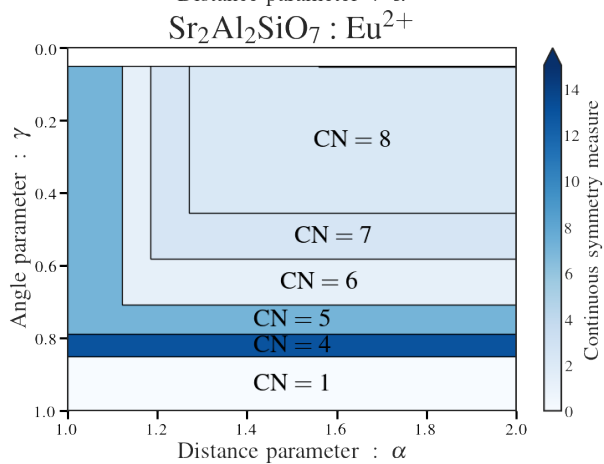
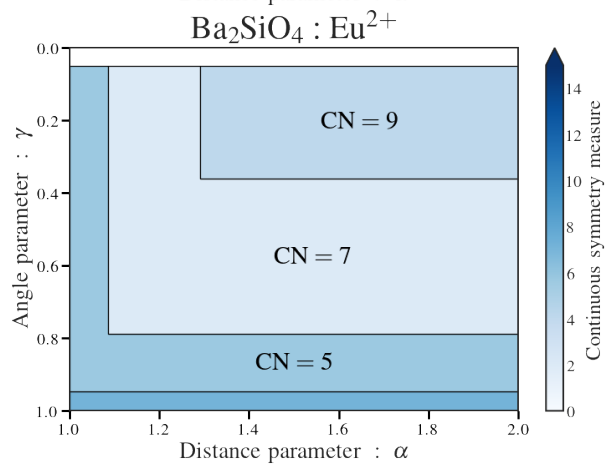

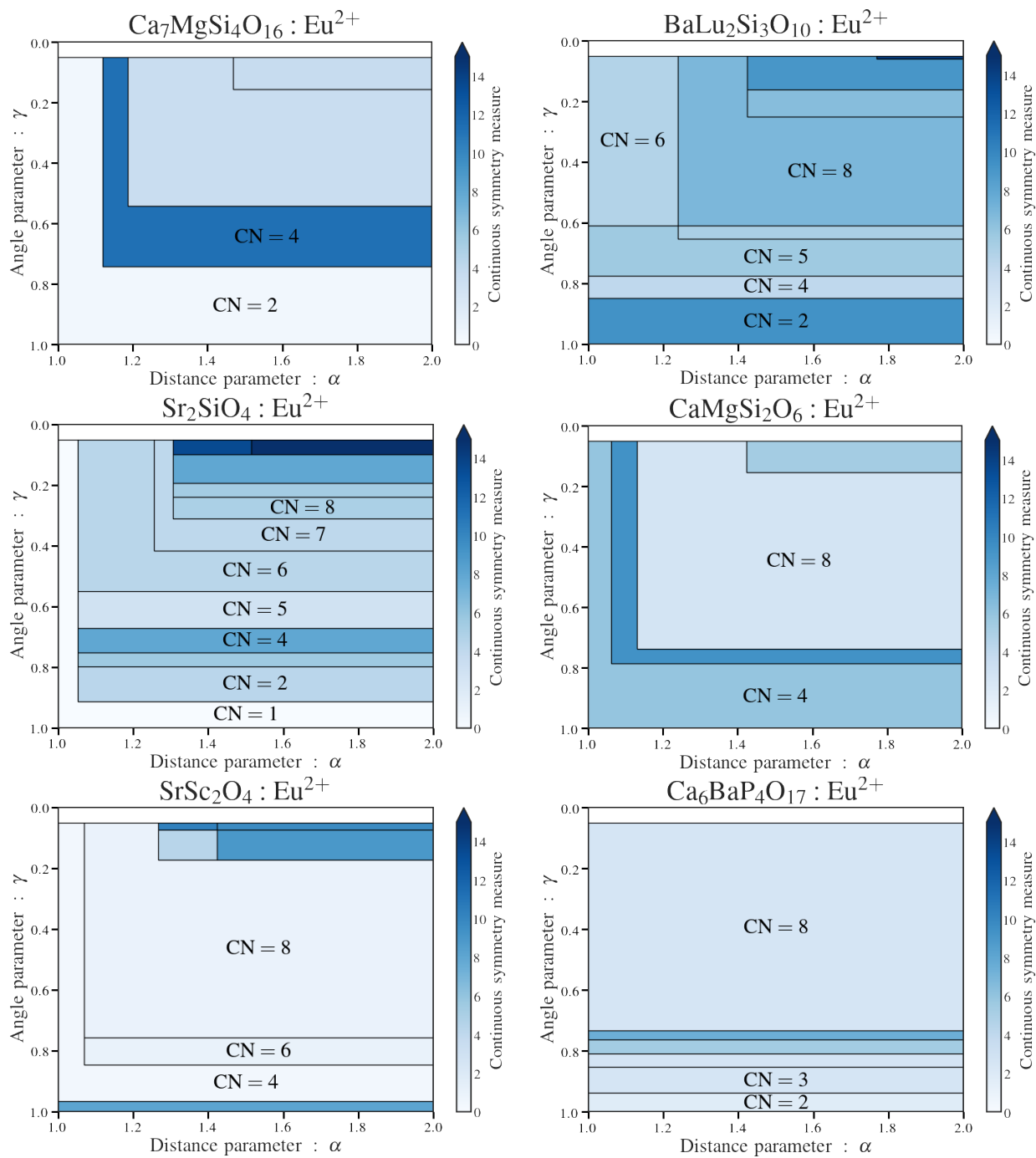

**Figure S7.** The Voronoi grid representation of $Eu^{2+}$ local environment of all $Eu^{2+}$-activated hosts in Table S1.